\documentclass[10pt,twocolumn,letterpaper]{article}

\usepackage[pagenumbers]{cvpr} 

\usepackage{graphicx}
\usepackage{amsmath}
\usepackage{amssymb}
\usepackage{booktabs}
\usepackage{amsfonts,amssymb}
\usepackage{multirow}
\usepackage{float}
\usepackage{caption}
\usepackage{subcaption}
\usepackage{fontawesome}
\usepackage{bbding}
\usepackage{tablefootnote}
\usepackage{wasysym}
\usepackage{threeparttable}
\usepackage[accsupp]{axessibility}
\usepackage[ruled,linesnumbered]{algorithm2e}
\usepackage{xcolor}
\usepackage{colortbl}

\def\ie{\textit{i.e.}}

\makeatletter
\renewcommand{\maketag@@@}[1]{\hbox{\m@th\normalsize\normalfont#1}}%
\makeatother

\usepackage[pagebackref,breaklinks,colorlinks]{hyperref}

\usepackage[capitalize]{cleveref}
\crefname{section}{Sec.}{Secs.}
\Crefname{section}{Section}{Sections}
\Crefname{table}{Table}{Tables}
\crefname{table}{Tab.}{Tabs.}

\def\cvprPaperID{12429} 
\def\confName{CVPR}
\def\confYear{2023}

\begin{document}

\title{Detecting Backdoors During the Inference Stage Based on Corruption Robustness Consistency}
\author{Xiaogeng Liu\textsuperscript{\rm 1, 4, 5, 6, 7}, Minghui Li\textsuperscript{\rm 3}, Haoyu Wang\textsuperscript{\rm 1, 4, 5, 6, 7},  Shengshan Hu\textsuperscript{\rm 1, 4, 5, 6, 7}\\ Dengpan Ye\textsuperscript{\rm 9}, Hai Jin\textsuperscript{\rm 2, 4, 5, 8}, Libing Wu\textsuperscript{\rm 9}, Chaowei Xiao\textsuperscript{\rm 10}\\
\textsuperscript{\rm 1}School of Cyber Science and Engineering, Huazhong University of Science and Technology\\
\textsuperscript{\rm 2}School of Computer Science and Technology, Huazhong University of Science and Technology\\
\textsuperscript{\rm 3}School of Software Engineering, Huazhong University of Science and Technology\\
\textsuperscript{\rm 4}National Engineering Research Center for Big Data Technology and System\\
\textsuperscript{\rm 5}Services Computing Technology and System Lab\\
\textsuperscript{\rm 6}Hubei Key Laboratory of Distributed System Security\\
\textsuperscript{\rm 7}Hubei Engineering Research Center on Big Data Security\quad
\textsuperscript{\rm 8}Cluster and Grid Computing Lab\\
\textsuperscript{\rm 9}School of Cyber Science and Engineering, Wuhan University\quad
\textsuperscript{\rm 10}Arizona State University\\
{\tt\small \{liuxiaogeng, minghuili, wwwwhy, hushengshan, hjin\}@hust.edu.cn}\\ 
{\tt\small \{yedp, wu\}@whu.edu.cn}\quad 
{\tt\small xiaocw@asu.edu}
}
\maketitle

\begin{abstract}
    Deep neural networks are proven to be vulnerable to backdoor attacks. Detecting the trigger samples during the inference stage, \ie, the test-time trigger sample detection, can prevent the backdoor from being triggered. However, existing detection methods often require the defenders to have high accessibility to victim models, extra clean data, or knowledge about the appearance of backdoor triggers, limiting their practicality. 
    
    In this paper, we propose the \underline{te}st-time \underline{co}rruption robustness consistency evaluation (TeCo)\footnote{\url{https://github.com/CGCL-codes/TeCo}}, a novel test-time trigger sample detection method that only needs the hard-label outputs of the victim models without any extra information. Our journey begins with the intriguing observation that the backdoor-infected models have similar performance across different image corruptions for the clean images, but perform discrepantly for the trigger samples. Based on this phenomenon, we design TeCo to evaluate test-time robustness consistency by calculating the deviation of severity that leads to predictions' transition across different corruptions. Extensive experiments demonstrate that compared with state-of-the-art defenses, which even require either certain information about the trigger types or accessibility of clean data, TeCo outperforms them on different backdoor attacks, datasets, and model architectures, enjoying a higher AUROC by $10\%$ and $5$ times of stability.
\end{abstract}

\section{Introduction}\label{intro}
Backdoor attacks have been shown to be a threat to \textit{deep neural networks} (DNNs)~\cite{DBLP:journals/access/GuLDG19,DBLP:conf/nips/NguyenT20,DBLP:conf/iccv/LiLWLHL21,souri2021sleeper}. A backdoor-infected DNN will perform normally on clean input data, 
but output the adversarially desirable target label when the input data are tampered with a special pattern (\ie, the backdoor trigger), which may cause serious safety issues. 

A critical dependency of a successful backdoor attack is that the attacker must provide the samples with backdoor triggers (we call them trigger samples for short hereafter)  to the infected models on the inference stage, otherwise, the backdoor will not be triggered. Thus, one way to counter the backdoor attacks is to judge whether the test data have triggers on it, \ie, the \textit{test-time trigger sample detection} (TTSD) defense\footnote{Some paper also call it online backdoor defense~\cite{ma2022beatrix,tang2021demon}.}~\cite{gao2019strip,udeshi2022model,chou2018sentinet}. This kind of defense can work corporately with other backdoor defenses such as model diagnosis defense~\cite{DBLP:conf/iclr/GuoLL22,xu2021detecting,dong2021black} or trigger reverse engineering~\cite{wang2019neural,tao2022better}, and also provide prior knowledge of the trigger samples in a comprehensive defense pipeline, which can help the down-steam defenses to statistically analyze the backdoor samples and mitigate the backdoor more effectively. 

On the other hand, the TTSD method, especially the black-box TTSD method can also serve as the last line of defense when someone adopts models with unknown credibility and has no authority to get access to the training data or model parameters, this scenario exists widely in the prevailing \textit{machine-learning-as-a-service} (MLaaS)~\cite{ribeiro2015mlaas,DBLP:journals/popets/HesamifardTGW18}.

However, with the development of backdoor attacks, the TTSD defense is facing great challenges. One of the major problems is that different types of triggers have been presented. Unlike the early backdoor attacks whose triggers are universal~\cite{DBLP:journals/access/GuLDG19,chen2017targeted} for all the images and usually conspicuous to human observers, recent works introduced sample-specific triggers~\cite{DBLP:conf/nips/NguyenT20} and even invisible triggers~\cite{DBLP:conf/iccv/LiLWLHL21,doan2021lira,DBLP:conf/iclr/NguyenT21,zhao2022defeat,DBLP:conf/mm/HuZZZZH022}, making it harder to apply pattern statistics or identify out-liners in the image space. Another main problem is the hardship of accomplishing the TTSD defense without extra knowledge such as supplemental data or model accessibility. On the other hand, existing TTSD methods require certain knowledge and assumption. Such assumptions include that the trigger is a specific type~\cite{udeshi2022model,gao2019strip}, the defenders have white-box accessibility to victim models, the predicted soft confidence score of each class~\cite{chou2018sentinet,gao2019strip} or  extra clean data for statistical analysis~\cite{zeng2021rethinking}, limiting the practicality for real-world applications. 

In this paper, we aim to design a TTSD defense free from these limitations. Specifically, we concentrate on a more practicable black-box hard-label backdoor setting~\cite{DBLP:conf/iclr/GuoLL22} where defenders can only get the final decision from the black-box victim models. In addition, no extra data is accessible and no assumption on trigger appearance is allowed. This setting assumes the defenders' ability as weak as possible and makes TTSD hard to achieve. To the best of our knowledge, we are the first to focus on the effectiveness of TTSD in this strict setting, and we believe it is desirable to develop TTSD methods working on such a scenario because it is very relevant to the wide deployment of cloud AI service~\cite{fowers2018configurable,chen2019cloud} and embedded AI devices~\cite{branco2019machine}. 

Since the setting we mentioned above has restricted the accessibility of victim models and the use of extra data, we cannot analyze the information in feature space~\cite{ma2022beatrix,tang2021demon} or train a trigger sample detector~\cite{zeng2021rethinking,DBLP:conf/iclr/DuJS20} like existing works. Fortunately, we find that the backdoor-infected models will present clearly different corruption robustness for trigger samples influenced by different image corruptions, but have relatively similar robustness throughout different image corruptions for clean samples, leaving the clue for trigger sample detection. We call these findings the \textit{anomalous corruption robustness consistency} of backdoor-infected models and describe them at length in Sec.~\ref{observation}. It is not the first time that image corruptions are  discussed in backdoor attacks and defenses~\cite{qiu2021deepsweep,li2020rethinking,liu2022adaptive}. However, previous works fail to explore the correlations between robustness against different corruptions, as discussed in Sec.~\ref{Difference}.

Based on our findings above, we propose \textit{\underline{te}st-time \underline{co}rruption robustness consistency evaluation} (TeCo), a novel test-time trigger sample detection method. At the inference stage of backdoor-infected models, TeCo modifies the input images by commonly used image corruptions~\cite{DBLP:conf/iclr/HendrycksD19} with growing severity and estimates the robustness against different types of corruptions from the hard-label outputs of the models. Then, a deviation measurement method is applied to calculate how spread out the results of robustness are. And TeCo makes the final judgment of whether the input images are with triggers based on this metric. 
Extensive experiments show that compared with the existing advanced TTSD method, TeCo improves AUROC about $10\%$, has a higher F1-score of $14\%$, and achieves $5$ times of stability against different types of trigger.

\begin{table}[t]
\setlength{\belowcaptionskip}{-0.5cm}
\centering
\resizebox{0.48\textwidth}{!}
{
\begin{tabular}{ccccccc}
    \toprule
    \multirow{2}[4]{*}{Method} & \multicolumn{2}{c}{Black-box Access} & No Need of  & \multicolumn{3}{c}{Trigger Aussmptions} \\
\cmidrule{2-3}\cmidrule{5-7}          & Logits-based & Decision-based & Clean Data & Universal & Sample-specific & Invisible \\
    \hline
    SentiNet~\cite{chou2018sentinet} &   \Circle    &   \Circle   &   \Circle    &   \CIRCLE    &    \Circle   &   \Circle\\
    SCan~\cite{tang2021demon} &   \Circle    &   \Circle   &   \Circle    &   \CIRCLE    &    \Circle   &   \Circle\\
    Beatrix~\cite{ma2022beatrix} &   \Circle    &   \Circle   &   \Circle    &   \CIRCLE    &    \CIRCLE   &   \CIRCLE\\
    \hline
    NEO\tablefootnote{NEO assumes the backdoor trigger is localized~\cite{DBLP:journals/access/GuLDG19} thus will be invalid on distributed or global triggers~\cite{chen2017targeted,zeng2021rethinking,DBLP:conf/nips/NguyenT20,doan2021lira}, including universal, sample-sepcific, and invisible ones.}~\cite{udeshi2022model} &   \CIRCLE    &   \CIRCLE   &   \CIRCLE    &   \Circle    &    \Circle    &   \Circle \\
    STRIP~\cite{gao2019strip} &   \CIRCLE    &   \Circle   &   \Circle    &   \CIRCLE    &    \Circle   &   \Circle\\
    FreqDetector~\cite{zeng2021rethinking} &   \CIRCLE    &   \CIRCLE   &   \Circle    &   \CIRCLE    &    \CIRCLE   &   \CIRCLE\\
    \textbf{TeCo (Ours)} &   \CIRCLE   &   \CIRCLE    &   \CIRCLE    &   \CIRCLE    &   \CIRCLE   &   \CIRCLE\\
    \bottomrule
\end{tabular}%
}
\caption{The model's accessibility, the use of clean data, and the assumptions on backdoor triggers required by various TTSD methods. We detail on some most related defenses in Sec.~\ref{relatedwork}. {"\CIRCLE"} represents the TTSD method supports this condition.
}
\label{related_works}%
\end{table}%

Finally, we take a deep investigation into our observations by constructing adaptive attacks against TeCo. From the results of feature space visualization and quantification of adaptive attacks, we speculate that the anomalous behavior of corruption robustness consistency derives from the widely-used dual-target training in backdoor attacks and it is hard to be avoided by existing trigger types. We hope these findings can shed light on a new perspective of backdoor attacks and defenses for the community.
In summary, we make the following contributions:
\vspace{-0.15cm}
\begin{itemize}
\item We propose TeCo, a novel test-time trigger sample detection method that only requires the hard-label outputs of the victim models and without extra data or assumptions about trigger types.
\vspace{-0.15cm}

\item We discover the fact of anomalous corruption robustness consistency, \ie, the backdoor-infected models have similar performance across different image corruptions for clean images, but not for the trigger samples. 
\vspace{-0.15cm}

\item We evaluate TeCo on five datasets, four model architectures (including CNNs and ViTs), and seven backdoor attacks with diverse trigger types. All experimental results support that TeCo outperforms  state-of-the-art methods.  
\vspace{-0.15cm}

\item We further analyze our observations by constructing adaptive attacks against TeCo. Experiments show that the widely-used dual-target training in backdoor attacks leads to anomalous corruption robustness consistency and it is hard to be avoided by existing backdoor triggers.
\end{itemize}

\section{Related Works}\label{relatedwork}
\subsection{Backdoor Attacks}
Badnets~\cite{DBLP:journals/access/GuLDG19} is the first work that describes how to embed a backdoor into the DNNs by poisoning part of the training data. Many backdoor attacks have been developed after Badnets. To categorize these works, a reasonable way is to divide them by the triggers' appearance of these attacks, \ie, the trigger types. The universal trigger is a classical trigger type that is leveraged in many works such as Badnets~\cite{DBLP:journals/access/GuLDG19}, Blended~\cite{chen2017targeted}, and Low-frequency~\cite{zeng2021rethinking}. Universal trigger represents that for any input images tampered with the same trigger, the backdoor-infected model will give the predefined predictions. Then, the sample-specific trigger is invented~\cite{DBLP:conf/nips/NguyenT20,DBLP:conf/iccv/LiLWLHL21,salem2022dynamic}. Unlike universal triggers, the appearance of sample-specific triggers depends on the images that the trigger attaches. Another research topic in backdoor attacks is the imperceptibility of the backdoor triggers, \ie, the invisible backdoor attacks, such as Wanet~\cite{DBLP:conf/iclr/NguyenT21}, LIRA~\cite{doan2021lira}, and SSBA~\cite{DBLP:conf/iccv/LiLWLHL21}. The triggers generated by this kind of attack only lead to subtle modifications in images, and humans can hardly perceive the existence of backdoor triggers. A backdoor attack can meet the sample-specific and invisible conditions simultaneously, for example, the SSBA attack~\cite{DBLP:conf/iccv/LiLWLHL21}.

\subsection{Backdoor Defenses}
In this paper, we mainly discuss the works which use trigger sample detection as a defense method. Since the success of backdoor attacks depends on the existence of trigger samples, detecting those trigger samples in training data or test data is a reasonable way to defend against backdoor attacks. Some detection methods focus on filtering trigger samples in the training stage and attempt to eliminate the backdoor attacks by preventing them from poisoning training data~\cite{tran2018spectral,pmlr-v139-hayase21a,chen2018detecting}. On the other hand, the \textit{test-time trigger sample detection} (TTSD) is developed since defenders cannot always control the training process of victim models. 

The first black-box TTSD method is STRIP~\cite{gao2019strip}. STRIP superimposes various clean images on the suspicious samples and evaluates the randomness of the model's logits outputs. NEO~\cite{udeshi2022model} assumes the backdoor trigger is localized and detects the trigger samples by masking random areas of the suspicious samples and repainting them with the dominant color. FreqDetector~\cite{zeng2021rethinking} finds that backdoor triggers often cause artifacts in the frequency space of the suspicious samples, and detects the trigger samples by training a frequency detector on clean images with data augmentations. Some other works assume they have white-box accessibility of the backdoor-infected models and detect the trigger samples by the salient maps~\cite{chou2018sentinet} or the features of intermediate layers~\cite{tang2021demon,ma2022beatrix}.

Since TTSD methods often make judgments based on certain statistical patterns of trigger samples, the advances in trigger types mentioned above put great threats to the TTSD methods with no doubt. As shown in Tab.~\ref{related_works}, we conclude that existing TTSD methods have relaxed their restrictions of defenders to achieve satisfying performance, leading to incomplete black-box settings, such as the requirement for model accessibility~\cite{chou2018sentinet,doan2020februus,tang2021demon,ma2022beatrix,jin2022can}, use of clean data~\cite{zeng2021rethinking,gao2019strip,do2022towards,liu2022adaptive,DBLP:conf/iclr/DuJS20}, and assumptions on specific trigger type~\cite{udeshi2022model,gao2019strip,tang2021demon,doan2020februus}.

\begin{figure*}[t]
\centering
    \begin{subfigure}{0.98\textwidth}
        \includegraphics[width=\textwidth]{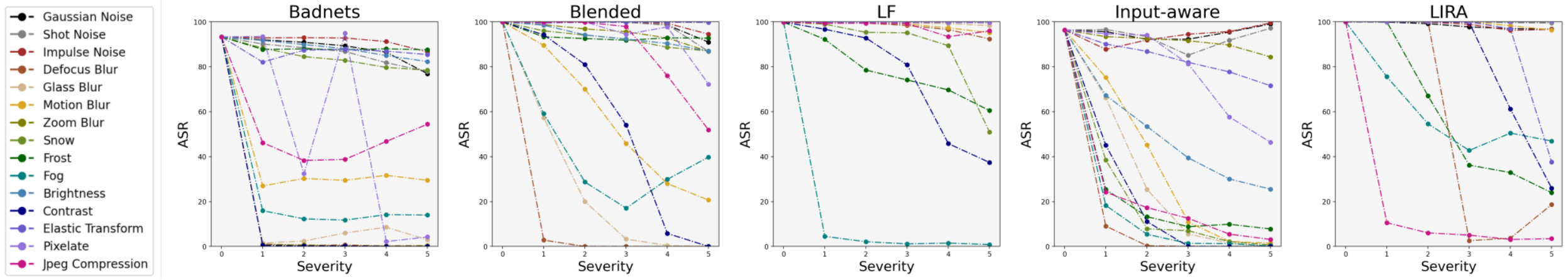} 
        \caption{The curves of ASR on trigger samples}
        \label{asr_sec3}
    \end{subfigure}
    \\
    \setlength{\belowcaptionskip}{-0.3cm}
    \begin{subfigure}{0.98\textwidth}
        \includegraphics[width=\textwidth]{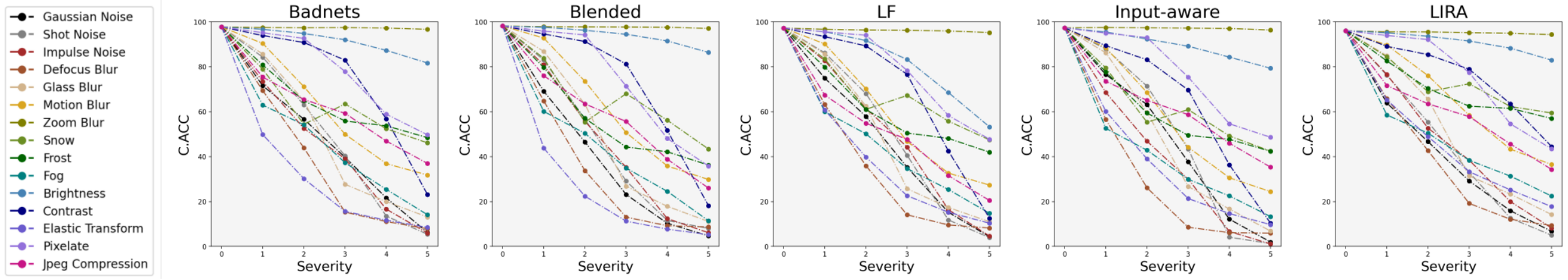} 
        \caption{The curves of ACC on clean images}
        \label{acc_sec3}
    \end{subfigure}
\setlength{\belowcaptionskip}{-0.3cm}
\caption{(a): The backdoor-infected model's \textit{attack success rate} (ASR) when trigger samples are tempered with different corruptions and levels of severity. (b): The accuracy (ACC) of clean images tempered with different corruptions and levels of severity. The curves separate loosely in (a), while the majority of curves gather more tightly in (b). This indicates that the backdoor-infected models have varied corruption robustness against different image corruptions on trigger samples, but have similar robustness against different image corruptions on clean samples.}
\label{observations_fig}
\vspace{-0.3cm}
\end{figure*}
\section{Corruption Robustness Consistency}\label{observation}
Before introducing our black-box trigger sample detection method, we first delineate the important findings that we discover from backdoor-infected models: \textit{given a backdoor-infected model, it will show clearly different robustness for trigger samples influenced by different image corruptions. However, for the clean images, the model will show similar robustness against the majority of image corruptions.} We stress that these phenomena exist widely in different backdoor-infected models.

\subsection{Corruption Robustness Consistency Test}\label{3.1}
We gain our findings by conducting \textit{Corruption Robustness Consistency} (CRC) test on backdoor-infected models. Given an infected model $C_\theta$, and an image corruption set ${\mathcal D}_K^N$ which has $K$ corruption types and $N$ levels of severity, CRC test computes the clean accuracy (ACC) of the clean images tempered with different image corruptions, or evaluates the \textit{attack success rate} (ASR) of the trigger samples tempered with different image corruptions. CRC test builds a list $L_{K,N}$ of ACC or ASR, where each element in this list is calculated by:
\begin{small}
\begin{equation}\label{eq:CRC_test}
   L_{k,d} = \left\{ \begin{aligned}
             &\frac{1}{I}\sum^I_{i=1}{\mathbb{I}}(C_\theta(D^k_n(x_i))=y_i),       \text{for clean samples}\\
             &\frac{1}{J}\sum^J_{j=1}{\mathbb{I}}(C_\theta(D^k_n(\hat{x_j}))=y_t), \text{for trigger samples}
             \end{aligned}
             \right.
\end{equation}
\end{small}where $I$ is the number of clean images, $J$ is the number of trigger samples, $x_i$ represents the clean image, $\hat{x_j}$ represents the trigger sample, and $D^k_n$ represents the $k$-th image corruption in the corruption set ${\mathcal D}_K^N$ with severity $n$. $y_i$ is the ground-truth label of $x_i$, $y_t$ is the target label that the adversaries want the infected model to predict when the trigger sample is given. $\mathbb{I}(\cdot)$ is an indicator function, where $I(A) = 1$ if and only if $A$ is true.

\subsection{Anomalous CRC of Backdoor-infected Models}
The list $L_{K,N}$ built in CRC test can be used to measure the corruption robustness of backdoor-infected models. We choose the image corruption set described in~\cite{DBLP:conf/iclr/HendrycksD19}, where the common image corruptions are categorized into $15$ classes and each kind of corruptions has $5$ levels of severity, then conduct CRC test on models infected by five backdoor attacks. From the visualization results in Fig.~\ref{observations_fig}(b), the majority of curves are relatively clustered and show a downward trend. We describe this phenomenon as the model has good corruption robustness consistency, because the model performs similarly on different image corruptions.

However, in Fig.~\ref{observations_fig}(a), the curves are more separated, indicating that the model has contrasting robustness against different image corruptions. Consequently, the model can be regarded as having bad corruption robustness consistency on trigger samples. Compared with the observations about Fig.~\ref{observations_fig}(b), the backdoor-infected models are suspicious to have different corruption robustness consistency on clean samples and trigger samples, \ie, the phenomenon of anomalous CRC. 

\begin{figure}[t]
\centering
\setlength{\belowcaptionskip}{-0.1cm}
    \begin{subfigure}{0.22\textwidth}
        \includegraphics[width=\textwidth]{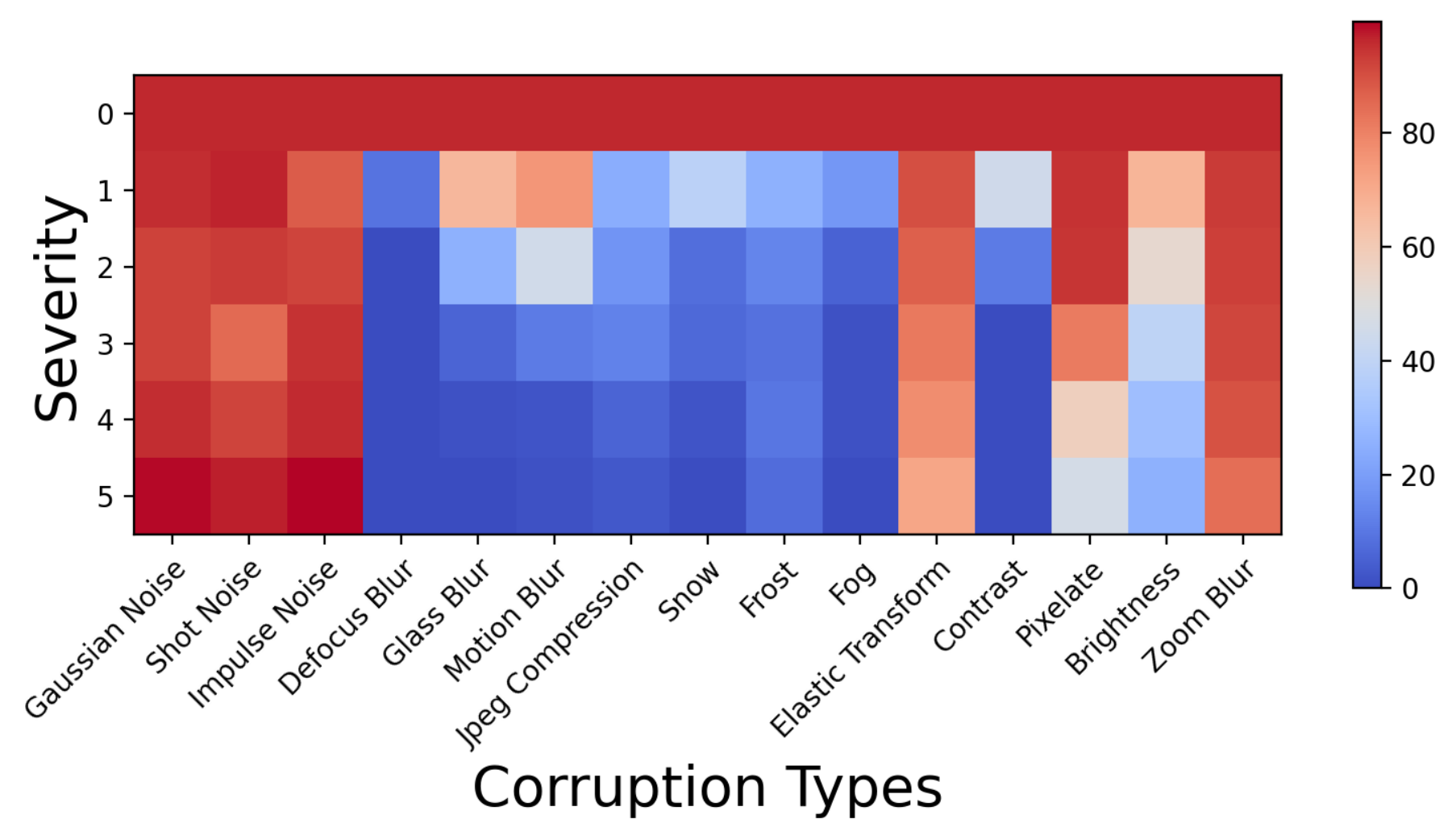} 
        \caption{ASR of trigger samples}
        \label{asr_heatmap_sec3}
    \end{subfigure}
    \begin{subfigure}{0.22\textwidth}
        \includegraphics[width=\textwidth]{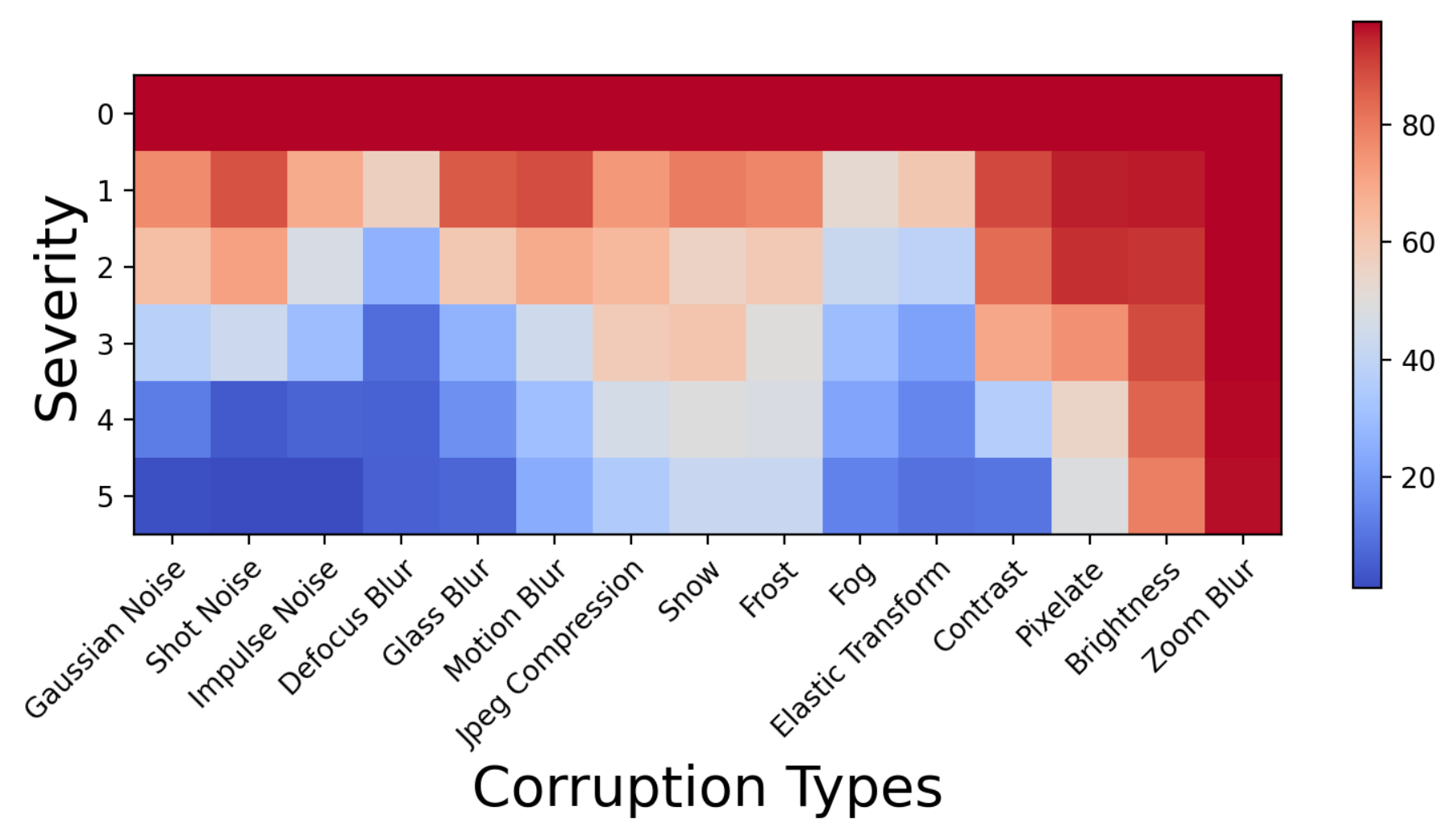}
        \caption{ACC of clean samples}
        \label{acc_heatmap_sec3}
    \end{subfigure}
\vspace{-0.2cm}
\caption{Take input-aware attack infected model as an example. Compared with clean samples, trigger samples have a more uneven heat map, which means that the backdoor-infected model are very robust on certain corruptions but also pretty vulnerable to some other corruptions.}
\vspace{-0.4cm}
\label{observations_heatmap}
\end{figure}

\subsection{Difference Between CRC and Previous Works}\label{Difference}
\vspace{-0.05cm}
Some previous works have discussed the corruption (or transformation) robustness of backdoor-infected models before~~\cite{qiu2021deepsweep,li2020rethinking,liu2022adaptive}. Specifically, Gaussian noise is investigated in~\cite{liu2022adaptive}, where the authors argue that adding this kind of noise can lead to abnormal behavior of backdoor-infected models on trigger samples. In~\cite{qiu2021deepsweep}, image transformations are used in a two-stage defense pipeline, where the defenders first fine-tune the infected models on one set of transformations and uses another set of transformations on the inference stage. This work is different from ours since it changes the parameters of backdoor-infected models, while we mainly focus on the characteristics of backdoor-infected models without modifications. In~\cite{li2020rethinking}, the authors evaluate the robustness of backdoor-infected models against multiple image transformations. They argue that some image transformations can mitigate the backdoor while others cannot, which is similar to our findings. But they still rely on a single transformation to defend against backdoor attacks and thus fail to leverage the difference of robustness.
\vspace{-0.1cm}
\section{Test-time CRC Evaluation (TeCo)}\label{method}
\vspace{-0.05cm}
In this section, we describe how we build our method based on the phenomenon of anomalous CRC.
\vspace{-0.05cm}

\subsection{Preliminaries}
The objective of backdoor attacks is to make the infected model behave normally on clean images but give predefined predictions on trigger samples. Thus, the target~\cite{li2022backdoor,DBLP:conf/iclr/GuoLL22} of backdoor attackers is training an infected model $C$ with parameters $\theta$ by:
\begin{small}
\begin{equation}\label{eq:backdoor_attack}
\begin{aligned}
\theta = \arg\min\limits_{\theta}&\mathbb{E}_{(x, y)\sim{\mathcal P}_S}{\mathcal J}(C(x;\theta),y) \\
+&\mathbb{E}_{(\hat{x}, y_t)\sim{\mathcal P}_{\hat{S}}}{\mathcal J}(C(\hat{x};\theta),y_t),
\end{aligned}
\end{equation}
\end{small}where $S$, $\hat{S}$ represent the clean data and trigger data, respectively, and $\mathcal J$ is the loss function.

TTSD methods work on a trained backdoor-infected model $C_\theta$ and a test dataset which contains clean samples and trigger samples ${{\mathcal T}=\{T\cup\hat{T}\}}$. The goal of designing TTSD is to find a method $M$:
\begin{small}
\begin{equation}\label{eq:backdoor_detection}
\begin{aligned}
M = \arg\max\limits_{M}&\mathbb{E}_{(x)\sim{\mathcal P}_T}{\mathbb{I}}(M(x,C_\theta)=0) \\
+&\mathbb{E}_{(\hat{x})\sim{\mathcal P}_{\hat{T}}}{\mathbb{I}}(M(\hat{x},C_\theta)=1).
\end{aligned}
\end{equation}
\end{small}

\subsection{Test-time CRC Evaluation}
To achieve Eq.~(\ref{eq:backdoor_attack}), the trigger sample detection methods should leverage contrasting characteristics of clean images and trigger samples. We have revealed in Sec.~\ref{observation} that backdoor-infected models have anomalous corruption robustness consistency, which is supported by the ACC and ASR evaluated from the entire test dataset. However, a question is how we can measure this property in test-time based on single input data.

A reasonable understanding is that the reduction of ACC or ASR is equivalent to the transitions of prediction labels. For example, if a model loses its accuracy on clean images with Gaussian noise, it can be regarded as the prediction labels of these images have changed compared with the original images'. Consequently, we can evaluate the corruption robustness consistency in the inference stage by adding image corruptions with growing severity, and recording the severity when the model's hard-label prediction gets changed. For different image corruptions, if the recorded severity levels are very similar, we can extrapolate that the corruption robustness consistency on the input image is high. 

After recording the levels of severity, the final step is to measure their dispersion. It is not hard to find such a metric since simply calculating the standard deviation is already effective according to our experiments. Alg.~\ref{alg.1} describes the detailed algorithm. TeCo maps the input image $x$ to a linearly separable space, and defenders can make judgments by a threshold $\gamma$:
\begin{small}
\begin{equation}\label{eq:make_judgements}
\begin{aligned}
      \Gamma(TeCo(x)) = 
      \left\{ \begin{aligned}
        &1 , TeCo(x) > \gamma\\
        &0 , TeCo(x) \leq \gamma
       \end{aligned}
 \right.
\end{aligned}
\end{equation}
\end{small}

\begin{table*}[htbp]
\setlength{\belowcaptionskip}{-0.2cm}
  \resizebox{1.0\textwidth}{!}
  {
  \centering
    \begin{threeparttable}
    \begin{tabular}{ccccccccccccccccccccc}
    \toprule
    \multirow{2}[2]{*}{Dataset} & \multirow{2}[2]{*}{Model} & Attack→ & \multicolumn{2}{c}{Badnets~\cite{DBLP:journals/access/GuLDG19}} & \multicolumn{2}{c}{Blended~\cite{chen2017targeted}} & \multicolumn{2}{c}{LF~\cite{zeng2021rethinking}} & \multicolumn{2}{c}{Input-aware~\cite{DBLP:conf/nips/NguyenT20}} & \multicolumn{2}{c}{Wanet~\cite{DBLP:conf/iclr/NguyenT21}} & \multicolumn{2}{c}{LIRA~\cite{doan2021lira}} & \multicolumn{2}{c}{SSBA~\cite{DBLP:conf/iccv/LiLWLHL21}} & \multicolumn{2}{c}{AVG(↑)} & \multicolumn{2}{c}{STD(↓)} \\
          &       & Detection↓ & AUROC & F1 score & AUROC & F1 score & AUROC & F1 score & AUROC & F1 score & AUROC & F1 score & AUROC & F1 score & AUROC & F1 score & AUROC & F1 score & AUROC & F1 score \\
    \midrule
    \multirow{6}[4]{*}{CIFAR10} & \multirow{3}[2]{*}{PreActResNet18} & STRIP & 0.790  & 0.743  & 0.726  & 0.685  & 0.973  & 0.937  & 0.283  & 0.526  & 0.395  & 0.526  & 0.555  & 0.661  & 0.364  & 0.526  & 0.584  & 0.658  & 0.236  & 0.140  \\
          &       & FreqDetector & 0.989  & 0.955  & 0.966  & 0.904  & 0.886  & 0.809  & 1.000  & 0.993  & 0.566  & 0.550  & 0.912  & 0.840  & 0.896  & 0.824  & 0.888  & 0.839  & 0.138  & 0.134  \\
          &       & \cellcolor[rgb]{ .906,  .902,  .902}Ours & \cellcolor[rgb]{ .906,  .902,  .902}0.911  & \cellcolor[rgb]{ .906,  .902,  .902}0.917  & \cellcolor[rgb]{ .906,  .902,  .902}0.935  & \cellcolor[rgb]{ .906,  .902,  .902}0.946  & \cellcolor[rgb]{ .906,  .902,  .902}0.939  & \cellcolor[rgb]{ .906,  .902,  .902}0.937  & \cellcolor[rgb]{ .906,  .902,  .902}0.905  & \cellcolor[rgb]{ .906,  .902,  .902}0.921  & \cellcolor[rgb]{ .906,  .902,  .902}0.915  & \cellcolor[rgb]{ .906,  .902,  .902}0.905  & \cellcolor[rgb]{ .906,  .902,  .902}0.953  & \cellcolor[rgb]{ .906,  .902,  .902}0.934  & \cellcolor[rgb]{ .906,  .902,  .902}0.868  & \cellcolor[rgb]{ .906,  .902,  .902}0.883  & \cellcolor[rgb]{ .906,  .902,  .902}\textbf{0.918 } & \cellcolor[rgb]{ .906,  .902,  .902}\textbf{0.920 } & \cellcolor[rgb]{ .906,  .902,  .902}\textbf{0.026 } & \cellcolor[rgb]{ .906,  .902,  .902}\textbf{0.020 } \\
\cmidrule{2-21}          & \multirow{3}[2]{*}{MobileViT-xs} & STRIP & 0.736  & 0.710  & 0.533  & 0.549  & 0.912  & 0.859  & 0.390  & 0.526  & 0.460  & 0.526  & 0.465  & 0.592  & 0.379  & 0.526  & 0.554  & 0.613  & 0.184  & 0.118  \\
          &       & FreqDetector & 0.989  & 0.955  & 0.966  & 0.904  & 0.834  & 0.763  & 0.996  & 0.972  & 0.510  & 0.526  & 0.980  & 0.940  & 0.896  & 0.824  & \textbf{0.882 } & 0.841  & 0.161  & 0.146  \\
          &       & \cellcolor[rgb]{ .906,  .902,  .902}Ours & \cellcolor[rgb]{ .906,  .902,  .902}0.682  & \cellcolor[rgb]{ .906,  .902,  .902}0.724  & \cellcolor[rgb]{ .906,  .902,  .902}0.927  & \cellcolor[rgb]{ .906,  .902,  .902}0.924  & \cellcolor[rgb]{ .906,  .902,  .902}0.917  & \cellcolor[rgb]{ .906,  .902,  .902}0.910  & \cellcolor[rgb]{ .906,  .902,  .902}0.811  & \cellcolor[rgb]{ .906,  .902,  .902}0.786  & \cellcolor[rgb]{ .906,  .902,  .902}0.913  & \cellcolor[rgb]{ .906,  .902,  .902}0.902  & \cellcolor[rgb]{ .906,  .902,  .902}0.964  & \cellcolor[rgb]{ .906,  .902,  .902}0.929  & \cellcolor[rgb]{ .906,  .902,  .902}0.920  & \cellcolor[rgb]{ .906,  .902,  .902}0.911  & \cellcolor[rgb]{ .906,  .902,  .902}0.876  & \cellcolor[rgb]{ .906,  .902,  .902}\textbf{0.870 } & \cellcolor[rgb]{ .906,  .902,  .902}\textbf{0.090 } & \cellcolor[rgb]{ .906,  .902,  .902}\textbf{0.075 } \\
    \midrule
    \multirow{6}[4]{*}{GTSRB} & \multirow{3}[2]{*}{PreActResNet18} & STRIP & 0.871  & 0.840  & 0.883  & 0.849  & 0.991  & 0.981  & 0.310  & 0.501  & 0.356  & 0.501  & 0.778  & 0.791  & 0.641  & 0.625  & 0.690  & 0.727  & 0.247  & 0.173  \\
          &       & FreqDetector & 0.981  & 0.939  & 0.993  & 0.960  & 0.964  & 0.901  & 0.925  & 0.848  & 0.483  & 0.503  & 0.595  & 0.562  & 0.544  & 0.548  & 0.784  & 0.752  & 0.213  & 0.188  \\
          &       & \cellcolor[rgb]{ .906,  .902,  .902}Ours & \cellcolor[rgb]{ .906,  .902,  .902}0.869  & \cellcolor[rgb]{ .906,  .902,  .902}0.835  & \cellcolor[rgb]{ .906,  .902,  .902}0.917  & \cellcolor[rgb]{ .906,  .902,  .902}0.913  & \cellcolor[rgb]{ .906,  .902,  .902}0.947  & \cellcolor[rgb]{ .906,  .902,  .902}0.962  & \cellcolor[rgb]{ .906,  .902,  .902}0.956  & \cellcolor[rgb]{ .906,  .902,  .902}0.959  & \cellcolor[rgb]{ .906,  .902,  .902}0.954  & \cellcolor[rgb]{ .906,  .902,  .902}0.961  & \cellcolor[rgb]{ .906,  .902,  .902}0.997  & \cellcolor[rgb]{ .906,  .902,  .902}0.986  & \cellcolor[rgb]{ .906,  .902,  .902}0.943  & \cellcolor[rgb]{ .906,  .902,  .902}0.967  & \cellcolor[rgb]{ .906,  .902,  .902}\textbf{0.940 } & \cellcolor[rgb]{ .906,  .902,  .902}\textbf{0.940 } & \cellcolor[rgb]{ .906,  .902,  .902}\textbf{0.036 } & \cellcolor[rgb]{ .906,  .902,  .902}\textbf{0.048 } \\
\cmidrule{2-21}          & \multirow{3}[2]{*}{MobileViT-xs} & STRIP & 0.947  & 0.939  & 0.875  & 0.856  & 0.962  & 0.937  & 0.285  & 0.501  & 0.438  & 0.501  & 0.616  & 0.687  & 0.544  & 0.552  & 0.667  & 0.710  & 0.246  & 0.184  \\
          &       & FreqDetector & 0.981  & 0.939  & 0.993  & 0.960  & 0.922  & 0.840  & 1.000  & 0.999  & 0.471  & 0.511  & 0.870  & 0.784  & 0.544  & 0.548  & 0.826  & 0.797  & 0.207  & 0.182  \\
          &       & \cellcolor[rgb]{ .906,  .902,  .902}Ours & \cellcolor[rgb]{ .906,  .902,  .902}0.914  & \cellcolor[rgb]{ .906,  .902,  .902}0.903  & \cellcolor[rgb]{ .906,  .902,  .902}0.924  & \cellcolor[rgb]{ .906,  .902,  .902}0.935  & \cellcolor[rgb]{ .906,  .902,  .902}0.993  & \cellcolor[rgb]{ .906,  .902,  .902}0.988  & \cellcolor[rgb]{ .906,  .902,  .902}0.847  & \cellcolor[rgb]{ .906,  .902,  .902}0.879  & \cellcolor[rgb]{ .906,  .902,  .902}0.973  & \cellcolor[rgb]{ .906,  .902,  .902}0.960  & \cellcolor[rgb]{ .906,  .902,  .902}0.987  & \cellcolor[rgb]{ .906,  .902,  .902}0.952  & \cellcolor[rgb]{ .906,  .902,  .902}0.939  & \cellcolor[rgb]{ .906,  .902,  .902}0.959  & \cellcolor[rgb]{ .906,  .902,  .902}\textbf{0.940 } & \cellcolor[rgb]{ .906,  .902,  .902}\textbf{0.939 } & \cellcolor[rgb]{ .906,  .902,  .902}\textbf{0.047 } & \cellcolor[rgb]{ .906,  .902,  .902}\textbf{0.034 } \\
    \midrule
    \multirow{6}[4]{*}{CIFAR100} & \multirow{3}[2]{*}{PreActResNet18} & STRIP & 0.860  & 0.812  & 0.769  & 0.719  & 0.955  & 0.898  & 0.249  & 0.502  & 0.485  & 0.503  & 0.589  & 0.590  & 0.685  & 0.651  & 0.656  & 0.668  & 0.221  & 0.140  \\
          &       & FreqDetector & 0.979  & 0.923  & 0.961  & 0.897  & 0.837  & 0.783  & 0.997  & 0.976  & 0.440  & 0.503  & 0.954  & 0.896  & 0.889  & 0.807  & 0.865  & 0.826  & 0.181  & 0.146  \\
          &       & \cellcolor[rgb]{ .906,  .902,  .902}Ours & \cellcolor[rgb]{ .906,  .902,  .902}0.939  & \cellcolor[rgb]{ .906,  .902,  .902}0.921  & \cellcolor[rgb]{ .906,  .902,  .902}0.939  & \cellcolor[rgb]{ .906,  .902,  .902}0.945  & \cellcolor[rgb]{ .906,  .902,  .902}0.834  & \cellcolor[rgb]{ .906,  .902,  .902}0.838  & \cellcolor[rgb]{ .906,  .902,  .902}0.878  & \cellcolor[rgb]{ .906,  .902,  .902}0.873  & \cellcolor[rgb]{ .906,  .902,  .902}0.971  & \cellcolor[rgb]{ .906,  .902,  .902}0.959  & \cellcolor[rgb]{ .906,  .902,  .902}0.913  & \cellcolor[rgb]{ .906,  .902,  .902}0.826  & \cellcolor[rgb]{ .906,  .902,  .902}0.968  & \cellcolor[rgb]{ .906,  .902,  .902}0.968  & \cellcolor[rgb]{ .906,  .902,  .902}\textbf{0.920 } & \cellcolor[rgb]{ .906,  .902,  .902}\textbf{0.904 } & \cellcolor[rgb]{ .906,  .902,  .902}\textbf{0.046 } & \cellcolor[rgb]{ .906,  .902,  .902}\textbf{0.054 } \\
\cmidrule{2-21}          & \multirow{3}[2]{*}{MobileViT-xs} & STRIP & 0.847  & 0.798  & 0.800  & 0.744  & 0.940  & 0.888  & 0.430  & 0.519  & 0.479  & 0.503  & 0.609  & 0.639  & 0.808  & 0.750  & 0.702  & 0.692  & 0.181  & 0.133  \\
          &       & FreqDetector & 0.979  & 0.923  & 0.961  & 0.897  & 0.914  & 0.838  & 0.999  & 0.990  & 0.426  & 0.503  & 0.941  & 0.877  & 0.889  & 0.807  & 0.873  & 0.834  & 0.186  & 0.146  \\
          &       & \cellcolor[rgb]{ .906,  .902,  .902}Ours & \cellcolor[rgb]{ .906,  .902,  .902}0.905  & \cellcolor[rgb]{ .906,  .902,  .902}0.909  & \cellcolor[rgb]{ .906,  .902,  .902}0.946  & \cellcolor[rgb]{ .906,  .902,  .902}0.957  & \cellcolor[rgb]{ .906,  .902,  .902}0.972  & \cellcolor[rgb]{ .906,  .902,  .902}0.967  & \cellcolor[rgb]{ .906,  .902,  .902}0.940  & \cellcolor[rgb]{ .906,  .902,  .902}0.932  & \cellcolor[rgb]{ .906,  .902,  .902}0.898  & \cellcolor[rgb]{ .906,  .902,  .902}0.881  & \cellcolor[rgb]{ .906,  .902,  .902}0.991  & \cellcolor[rgb]{ .906,  .902,  .902}0.965  & \cellcolor[rgb]{ .906,  .902,  .902}0.956  & \cellcolor[rgb]{ .906,  .902,  .902}0.955  & \cellcolor[rgb]{ .906,  .902,  .902}\textbf{0.944 } & \cellcolor[rgb]{ .906,  .902,  .902}\textbf{0.938 } & \cellcolor[rgb]{ .906,  .902,  .902}\textbf{0.031 } & \cellcolor[rgb]{ .906,  .902,  .902}\textbf{0.030 } \\
    \midrule
    \multirow{6}[4]{*}{Tiny-ImageNet} & \multirow{3}[2]{*}{PreActResNet18} & STRIP & 0.852  & 0.788  & 0.949  & 0.892  & 0.995  & 0.976  & 0.430  & 0.504  & 0.681  & 0.640  & 0.511  & 0.545  & 0.767  & 0.722  & 0.741  & 0.724  & 0.198  & 0.162  \\
          &       & FreqDetector & 0.710  & 0.652  & 0.999  & 0.989  & 0.920  & 0.828  & 1.000  & 0.996  & 0.655  & 0.617  & 0.960  & 0.910  & 0.992  & 0.958  & 0.891  & 0.850  & 0.135  & 0.147  \\
          &       & \cellcolor[rgb]{ .906,  .902,  .902}Ours & \cellcolor[rgb]{ .906,  .902,  .902}0.987  & \cellcolor[rgb]{ .906,  .902,  .902}0.982  & \cellcolor[rgb]{ .906,  .902,  .902}0.977  & \cellcolor[rgb]{ .906,  .902,  .902}0.978  & \cellcolor[rgb]{ .906,  .902,  .902}0.993  & \cellcolor[rgb]{ .906,  .902,  .902}0.989  & \cellcolor[rgb]{ .906,  .902,  .902}0.978  & \cellcolor[rgb]{ .906,  .902,  .902}0.974  & \cellcolor[rgb]{ .906,  .902,  .902}0.888  & \cellcolor[rgb]{ .906,  .902,  .902}0.889  & \cellcolor[rgb]{ .906,  .902,  .902}0.977  & \cellcolor[rgb]{ .906,  .902,  .902}0.974  & \cellcolor[rgb]{ .906,  .902,  .902}0.983  & \cellcolor[rgb]{ .906,  .902,  .902}0.977  & \cellcolor[rgb]{ .906,  .902,  .902}\textbf{0.969 } & \cellcolor[rgb]{ .906,  .902,  .902}\textbf{0.966 } & \cellcolor[rgb]{ .906,  .902,  .902}\textbf{0.033 } & \cellcolor[rgb]{ .906,  .902,  .902}\textbf{0.032 } \\
\cmidrule{2-21}          & \multirow{3}[2]{*}{MobileViT-xs} & STRIP & 0.737  & 0.688  & 0.872  & 0.809  & 0.991  & 0.964  & 0.421  & 0.516  & 0.647  & 0.615  & 0.585  & 0.655  & 0.766  & 0.716  & 0.717  & 0.709  & 0.174  & 0.134  \\
          &       & FreqDetector & 0.689  & 0.638  & 0.998  & 0.984  & 0.938  & 0.865  & 1.000  & 0.999  & 0.631  & 0.602  & 0.770  & 0.696  & 0.982  & 0.934  & 0.858  & 0.817  & 0.146  & 0.156  \\
          &       & \cellcolor[rgb]{ .906,  .902,  .902}Ours & \cellcolor[rgb]{ .906,  .902,  .902}0.979  & \cellcolor[rgb]{ .906,  .902,  .902}0.981  & \cellcolor[rgb]{ .906,  .902,  .902}0.974  & \cellcolor[rgb]{ .906,  .902,  .902}0.975  & \cellcolor[rgb]{ .906,  .902,  .902}0.982  & \cellcolor[rgb]{ .906,  .902,  .902}0.973  & \cellcolor[rgb]{ .906,  .902,  .902}0.984  & \cellcolor[rgb]{ .906,  .902,  .902}0.983  & \cellcolor[rgb]{ .906,  .902,  .902}0.986  & \cellcolor[rgb]{ .906,  .902,  .902}0.975  & \cellcolor[rgb]{ .906,  .902,  .902}0.938  & \cellcolor[rgb]{ .906,  .902,  .902}0.874  & \cellcolor[rgb]{ .906,  .902,  .902}0.975  & \cellcolor[rgb]{ .906,  .902,  .902}0.971  & \cellcolor[rgb]{ .906,  .902,  .902}\textbf{0.974 } & \cellcolor[rgb]{ .906,  .902,  .902}\textbf{0.962 } & \cellcolor[rgb]{ .906,  .902,  .902}\textbf{0.015 } & \cellcolor[rgb]{ .906,  .902,  .902}\textbf{0.036 } \\
    \midrule
    \multirow{6}[4]{*}{ImageNet200} & \multirow{3}[2]{*}{WideResNet101-2} & STRIP & 0.921  & 0.869  & 0.959  & 0.903  &  -    &  -    & 0.421  & 0.502  & 0.584  & 0.567  & 0.693  & 0.668  & 0.765  & 0.695  & 0.724  & 0.701  & 0.186  & 0.146  \\
          &       & FreqDetector & 0.526  & 0.520  & 0.998  & 0.986  &  -    &  -    & 1.000  & 1.000  & 0.484  & 0.517  & 0.979  & 0.949  & 0.994  & 0.970  & 0.830  & 0.824  & 0.230  & 0.216  \\
          &       & \cellcolor[rgb]{ .906,  .902,  .902}Ours & \cellcolor[rgb]{ .906,  .902,  .902}0.974  & \cellcolor[rgb]{ .906,  .902,  .902}0.979  & \cellcolor[rgb]{ .906,  .902,  .902}0.982  & \cellcolor[rgb]{ .906,  .902,  .902}0.983  & \cellcolor[rgb]{ .906,  .902,  .902} - & \cellcolor[rgb]{ .906,  .902,  .902} - & \cellcolor[rgb]{ .906,  .902,  .902}0.937  & \cellcolor[rgb]{ .906,  .902,  .902}0.920  & \cellcolor[rgb]{ .906,  .902,  .902}0.987  & \cellcolor[rgb]{ .906,  .902,  .902}0.977  & \cellcolor[rgb]{ .906,  .902,  .902}0.997  & \cellcolor[rgb]{ .906,  .902,  .902}0.996  & \cellcolor[rgb]{ .906,  .902,  .902}0.922  & \cellcolor[rgb]{ .906,  .902,  .902}0.938  & \cellcolor[rgb]{ .906,  .902,  .902}\textbf{0.966 } & \cellcolor[rgb]{ .906,  .902,  .902}\textbf{0.965 } & \cellcolor[rgb]{ .906,  .902,  .902}\textbf{0.027 } & \cellcolor[rgb]{ .906,  .902,  .902}\textbf{0.027 } \\
\cmidrule{2-21}          & \multirow{3}[2]{*}{SwinT-Base} & STRIP & 0.992  & 0.968  & 0.939  & 0.875  &  -    &  -    & 0.944  & 0.873  & 0.726  & 0.672  & 0.993  & 0.974  & 0.715  & 0.659  & 0.885  & 0.837  & 0.118  & 0.128  \\
          &       & FreqDetector & 0.526  & 0.520  & 0.998  & 0.986  &  -    &  -    & 1.000  & 1.000  & 0.455  & 0.504  & 0.974  & 0.940  & 0.994  & 0.970  & 0.825  & 0.820  & 0.237  & 0.218  \\
          &       & \cellcolor[rgb]{ .906,  .902,  .902}Ours & \cellcolor[rgb]{ .906,  .902,  .902}0.978  & \cellcolor[rgb]{ .906,  .902,  .902}0.978  & \cellcolor[rgb]{ .906,  .902,  .902}0.978  & \cellcolor[rgb]{ .906,  .902,  .902}0.979  & \cellcolor[rgb]{ .906,  .902,  .902} - & \cellcolor[rgb]{ .906,  .902,  .902} - & \cellcolor[rgb]{ .906,  .902,  .902}0.990  & \cellcolor[rgb]{ .906,  .902,  .902}0.988  & \cellcolor[rgb]{ .906,  .902,  .902}0.985  & \cellcolor[rgb]{ .906,  .902,  .902}0.970  & \cellcolor[rgb]{ .906,  .902,  .902}0.999  & \cellcolor[rgb]{ .906,  .902,  .902}0.998  & \cellcolor[rgb]{ .906,  .902,  .902}0.980  & \cellcolor[rgb]{ .906,  .902,  .902}0.975  & \cellcolor[rgb]{ .906,  .902,  .902}\textbf{0.985 } & \cellcolor[rgb]{ .906,  .902,  .902}\textbf{0.981 } & \cellcolor[rgb]{ .906,  .902,  .902}\textbf{0.007 } & \cellcolor[rgb]{ .906,  .902,  .902}\textbf{0.009 } \\
    \bottomrule
    \end{tabular}%
    \begin{tablenotes}
    \Large
    \item[*] LF is computationally infeasible on ImageNet200.
    \vspace{-0.4cm}
    \end{tablenotes}
    \end{threeparttable}
    }
\caption{The evaluation results on different attacks, datasets, and backbones. The last two columns show the average performance and the standard deviation of performance across different attacks. The best results are in bold. We highlight that our method not only has good effectiveness, but also keeps outstanding stability (about $5$ times of the runner-ups' on average) against different backdoor attacks including universal, sample-specific, and invisible ones.} 
\label{main_table}%
\vspace{-0.3cm}
\end{table*}%

\section{Experiments}\label{experiemnt}
\subsection{Experimental Settings} 
\textbf{Implementation details. } We take the common image corruptions introduced in~\cite{DBLP:conf/iclr/HendrycksD19} as the image corruption set $D_k^N$ in Alg.~\ref{alg.1}. This corruption set has $15$ diverse image corruptions with the severity ranging from $1$ to $5$. We choose standard deviation as the deviation measurement method\footnote{We investigate the choice of image corruption set and deviation measurement method in the supplementary.}.

\textbf{Attack methods. }We evaluate our method against seven backdoor attacks, including Badnets attack~\cite{DBLP:journals/access/GuLDG19}, Blended attack~\cite{chen2017targeted}, \textit{Low-frequency} (LF) attack~\cite{zeng2021rethinking}, Input-aware attack~\cite{DBLP:conf/nips/NguyenT20}, Wanet attack~\cite{DBLP:conf/iclr/NguyenT21}, LIRA attack~\cite{doan2021lira}, and SSBA attack~\cite{DBLP:conf/iccv/LiLWLHL21}. We follow an open-sourced backdoor benchmark~\cite{wu2022backdoorbench} for the training settings of these attacks. To ensure the attacks' strength, $10\%$ of the training set are poisoned. As illustrated in Tab.~\ref{attacks_type}, the attacks in our experiments contain different trigger types.

\begin{algorithm}[t]
\caption{Test-time CRC Evaluation (TeCo)}
\label{alg.1}
\KwIn{Test sample $x$; test model $C_\theta$; deviation measurement method $Dev$;image corruption set ${\mathcal D}_K^N$, where $K$ is the number of corruption types, and $N$ is the maximum of severity.}
\KwOut{Prediction score of test sample $x$.}
Initialize $\mathcal L \gets \{\}$, $P_{org} \gets C_\theta(x)$\;
\For{$k = 1$ to $K$}{
    $L \gets N+1$\;
    \For{$n = 1$ to $N$}{
        \If{$ C_\theta(D_k^n(x)) \neq P_{org}$}{
            $L \gets n$\;
            \textbf{break}\;
        }
    }
    $\mathcal L \gets \mathcal L \cup \{L\}$\;
}
$deviation \gets Dev(\mathcal L)$\;
\textbf{return} $deviation$
\end{algorithm}

\begin{table}[htbp]
\setlength{\belowcaptionskip}{-0.5cm}
\centering
\resizebox{0.48\textwidth}{!}
  {
    \begin{tabular}{c|ccccccc}
    \hline
    Types↓ Attacks→ & Badnets & Blended & LF & Input-aware & Wanet & LIRA  & SSBA \\
    \hline
    Universal & \Checkmark & \Checkmark & \Checkmark &       &       &       &  \\
    Sample-specific &       &       &       & \Checkmark &       & \Checkmark & \Checkmark \\
    Invisible &       &       &       &       & \Checkmark & \Checkmark & \Checkmark \\
    \hline
    \end{tabular}%
    }
\vspace{-0.2cm}
\caption{The backdoor attacks involved in our evaluations have covered the majority of trigger types.} 
\label{attacks_type}%
\end{table}%

\textbf{Datasets and backbones. }Five datasets and four backbones are involved in our experiments. The datasets include CIFAR10~\cite{krizhevsky2009learning}, CIFAR100~\cite{krizhevsky2009learning}, GTSRB~\cite{houben2013detection}, Tiny-ImageNet~\cite{le2015tiny}, and ImageNet200~\cite{russakovsky2015imagenet} which is used in~\cite{DBLP:conf/iccv/LiLWLHL21}. For images in relatively low size, we use PreActResNet18~\cite{he2016identity} and MobileViT-xs~\cite{DBLP:conf/iclr/MehtaR22} as the backbones. And for ImageNet200, we use WideResNet101-2~\cite{DBLP:conf/bmvc/ZagoruykoK16} and SwinTransformer-Base~\cite{liu2021swin}, and fine-tune them from checkpoints~\cite{rw2019timm} pre-trained on ImageNet1K. 

\textbf{Competitors. }Since TeCo is the first test-time trigger sample detection method that works in hard-label black-box settings and has no extra dependency, We compare our method with two trigger sample detection methods that work in looser conditions but still meet the black-box requirement. To the best of our knowledge, STRIP~\cite{gao2019strip} is the first black-box TTSD method, and it still serves as a baseline in many recent works~\cite{zeng2021rethinking,ma2022beatrix}; Frequency detector (FreqDetector)~\cite{zeng2021rethinking} is the state-of-the-art trigger sample detection method. We implement them following their official codes. STRIP needs the logits-based black-box accessibility, while FreqDetector has no requirement for accessing the backdoor models. In addition, both of them need extra clean data to accomplish the trigger detection task. 

\textbf{Evaluation metrics. }Two metrics are used: (1) The \textit{Area Under Receiver Operating Curve} (AUROC), which is a widely-used metric to measure the trade-off between the false positive rate for clean samples and true positive rate for trigger samples for a detection method. (2) The F1 score. We calculate the best F1 score of detection methods to evaluate their optimal performance. The F1 score in our experiments is computed by:
\begin{small}
\begin{equation}\label{eq:F1}
\begin{aligned}
\text{ F$1$ score} = \max\limits_{\gamma\in\Gamma} \frac{2 \times (\text{precision}_\gamma \times \text{recall}_\gamma)}{(\text{precision}_\gamma + \text{recall}_\gamma)},
\end{aligned}
\end{equation}
\end{small}where $\Gamma$ represents all possible thresholds.

We also include additional metrics such as FAR, FRR, and \textit{Backdoored Data Rejection Rate} (BDR). For more implementation details and evaluations, please also refer to the supplementary. 

\subsection{Effectiveness Studies}
We first evaluate the performance of TeCo on different backdoor-infected models comprehensively. As shown in Tab.~\ref{main_table}, TeCo can precisely detect the trigger samples in the inference stage with the average AUROC $\ge0.876$ for different backdoor attacks on diverse datasets and backbones. In addition, since in the real-world scenario, the TTSD methods should have solid effectiveness against different types of backdoor triggers, we investigate the stability of our method by calculating the standard deviations of its performance on different backdoor attacks with the same dataset and backbone. We highlight that TeCo achieves overall average AUROC $=0.9433$ and F1 score $=0.9386$, with the standard deviation of AUROCs and F1 scores equal to $0.0360$ and $0.0364$ respectively. These results indicate that TeCo outperforms the runner-up by about $10\%$ in terms of AUROC, $14\%$ in terms of F1-score, and achieves $5$ times of stability against different types of trigger.
In summary, our work maintains stable effectiveness among different trigger types, with only hard-label black-box model accessibility and no need for extra knowledge.

There are also some interesting results about baselines. Since the \textit{Low-frequency} (LF) attack is designed to avoid FreqDetector~\cite{zeng2021rethinking}, FreqDetector should have low effectiveness against this attack. However, we implement them following the official codes and find that if we let FreqDetector work in a binary classification manner and make judgments based on thresholds, it will perform well on LF attack. So we believe it is not unfair to involve LF attack and FreqDetector simultaneously in our experiments. Another interesting phenomenon is the success of STRIP against input-aware and LIRA attacks on SwinTransformer-base/ImageNet200. We further investigate it in our supplementary and show that the performance of STRIP is somehow influenced by the choice of backbones. 

\subsection{Ablation Studies}
\begin{figure}[t]
\setlength{\belowcaptionskip}{-0.3cm}
\centering
    \begin{subfigure}{0.15\textwidth}
        \includegraphics[width=\textwidth]{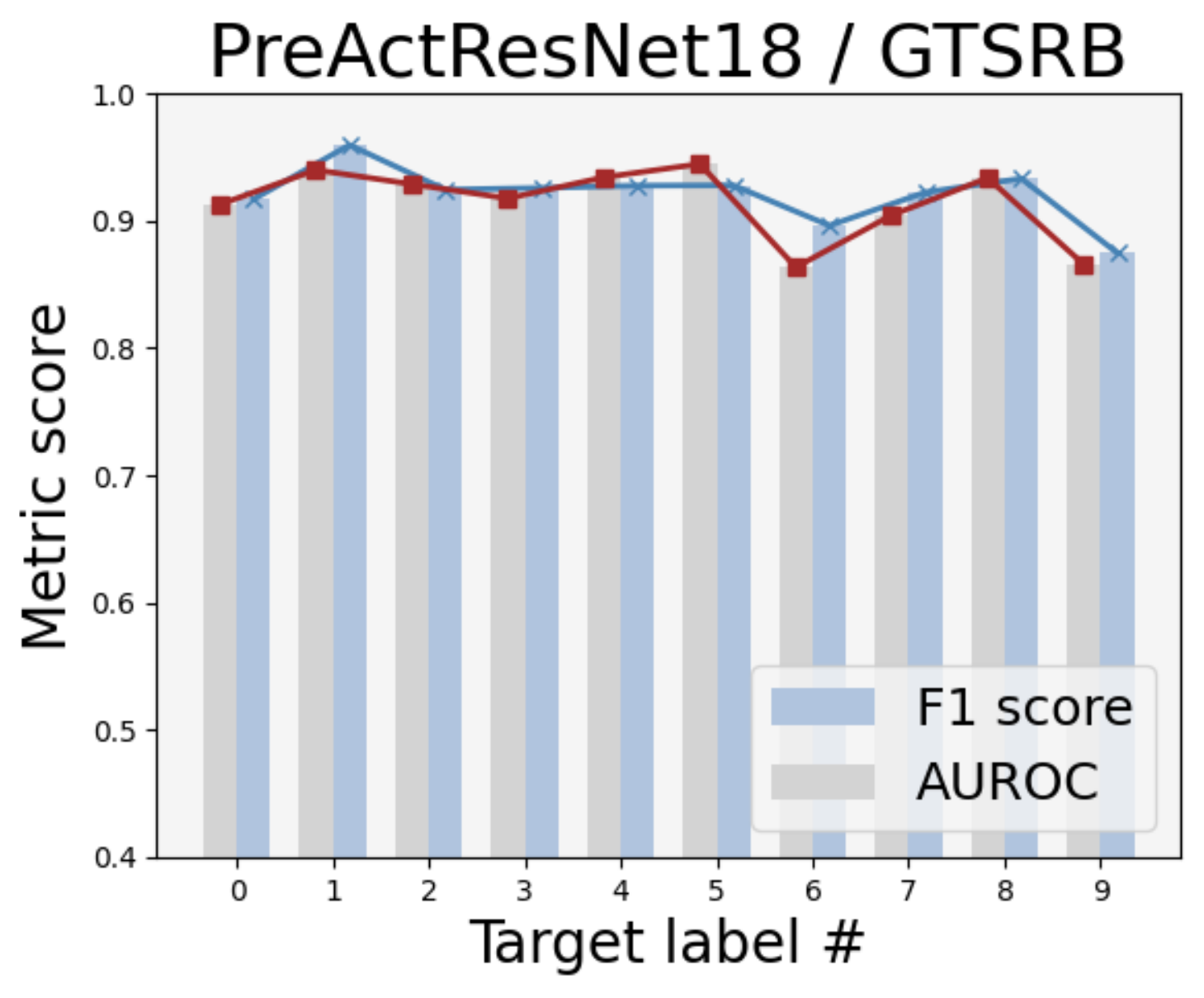} 
        \caption{Blended}
        \label{Fig.target_ours}
    \end{subfigure}
    \begin{subfigure}{0.15\textwidth}
        \includegraphics[width=\textwidth]{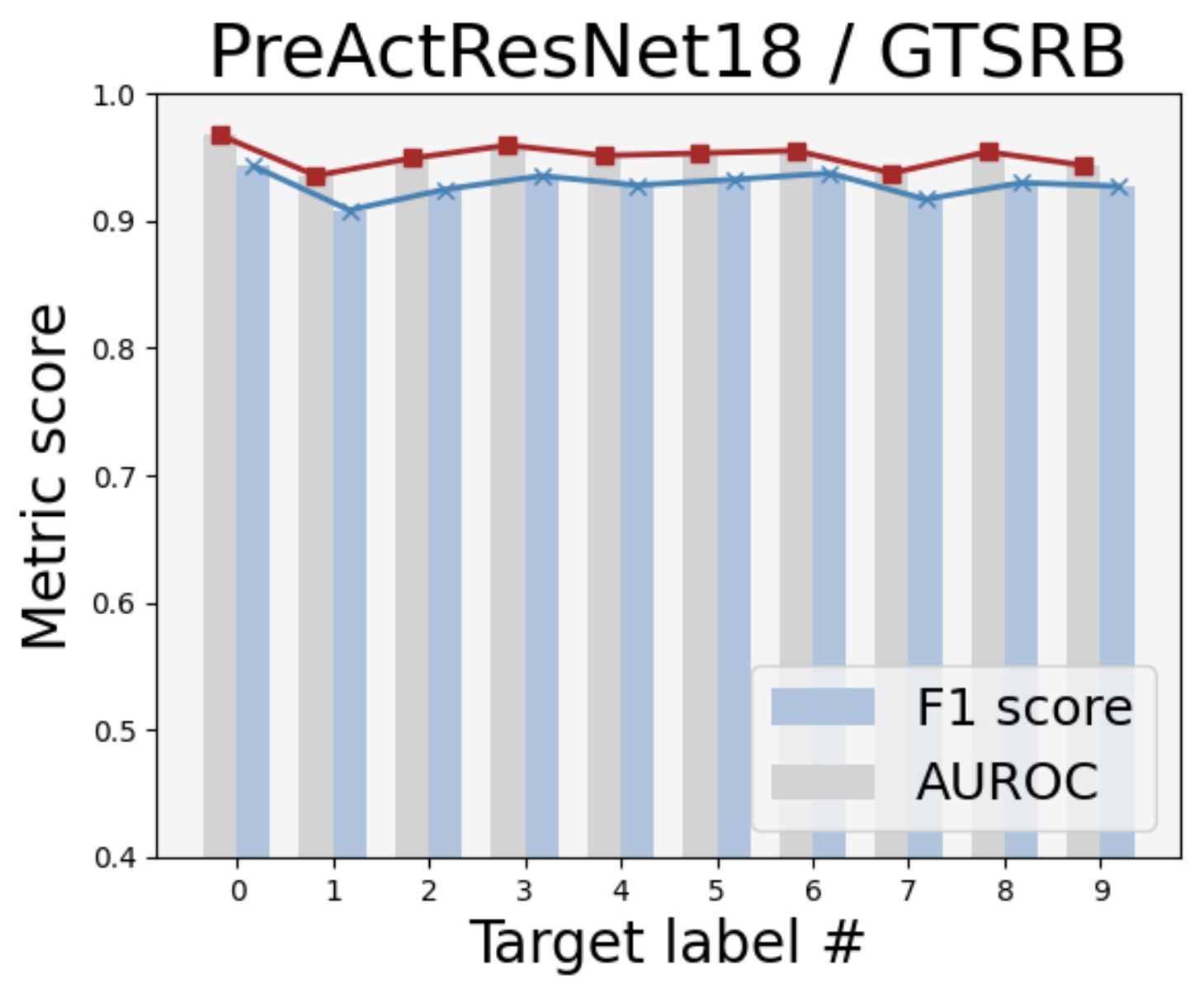} 
        \caption{SSBA}
        \label{Fig.target_strip}
    \end{subfigure}
    \begin{subfigure}{0.15\textwidth}
        \includegraphics[width=\textwidth]{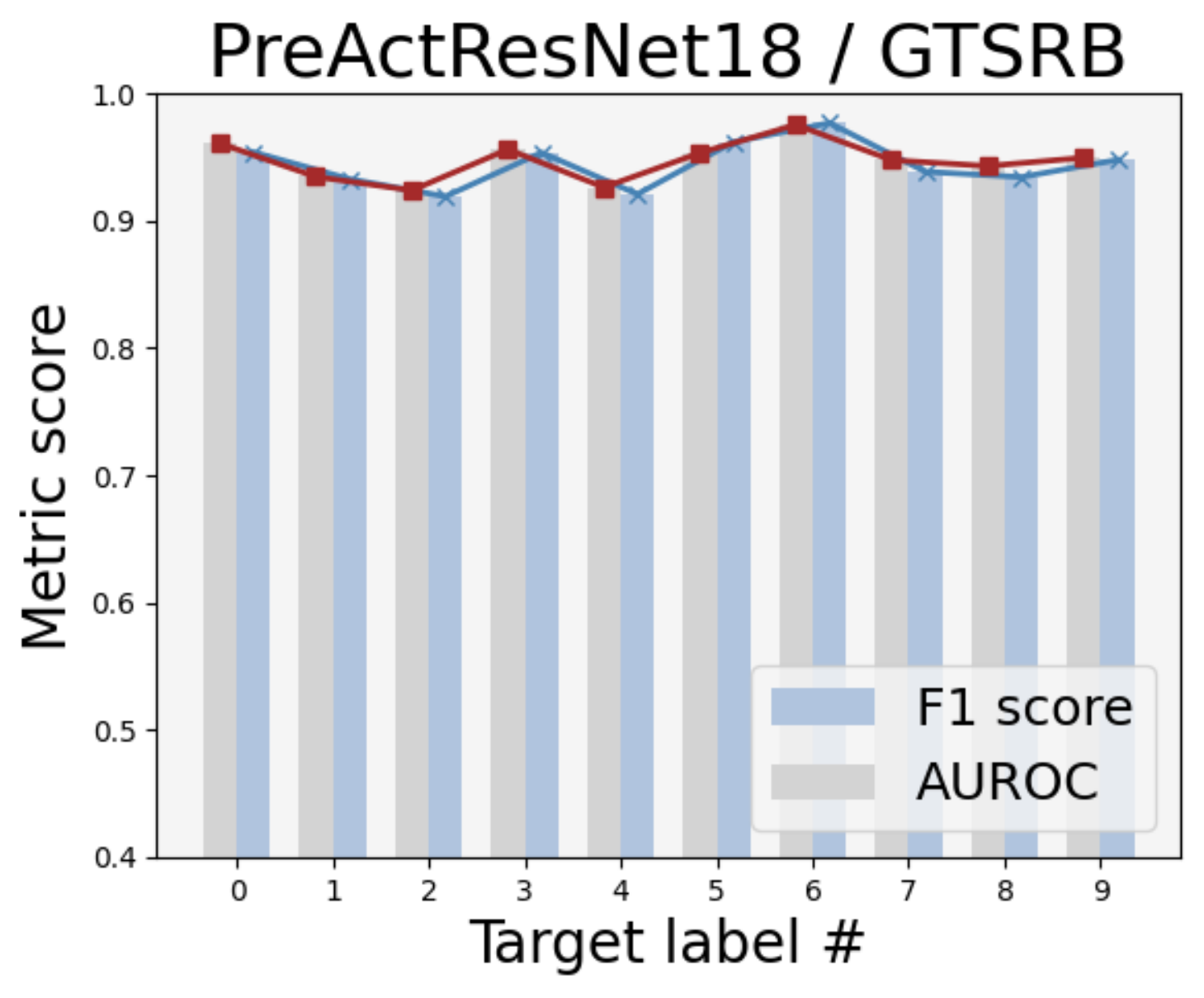} 
        \caption{Wanet}
        \label{Fig.target_detect}
    \end{subfigure}
\caption{Performance of TeCo against different target labels}
\vspace{-0.1cm}
\label{Fig.targets}
\end{figure}

\textbf{The impact of different target labels. }We further evaluate the effectiveness of TeCo against different predefined target labels. We select $3$ attacks (Blended, SSBA, and Wanet) from the seven attacks mentioned above to represent the universal, sample-specific, and invisible backdoor attacks, and make them attack $10$ different labels that we randomly select from GTSRB. Thus we have $30$ backdoor-infected models with different trojan labels and trigger types. Fig.~\ref{Fig.targets} illustrates the stability of TeCo against backdoor attacks with different target labels. For Blended, TeCo remains AUROC $\ge0.874$, and the \textit{standard deviation} (STD) of AUROC is about $0.021$. For SSBA, TeCo achieves AUROC $\ge0.908$ and STD $\approx0.010$. For Wanet, TeCo achieves AUROC $\ge0.919$ and STD $\approx0.017$. These results support that target trojan labels have little influence on TeCo's performance. 

\textbf{The impact of max severity $N$. }We mark the computational cost of vanilla inference process as $O_1(1)$ and the average computational cost of image corruptions as $O_2(1)$. The computational cost of TeCo in the worst case is $O_1(N\times K) + O_2(N\times K)$, where $K$ is the number of corruption types and $N$ is the maximum of severity. Thus, given fixed corruption types, the max severity has a critical influence on the running efficiency. Here, we investigate TeCo's performance with $N$ ranging from $1$ to $5$. As shown in Tab.~\ref{max_table}, TeCo maintains good effectiveness and stability in different $N$. We illustrate some results in Fig.~\ref{max_fig}, which shows that TeCo's performance grows with the increasing max severity $N$, and still gets satisfying effectiveness with low $N$. In other words, TeCo also has competitive effectiveness when the computational cost is limited.

\begin{table}[htbp]
\resizebox{0.47\textwidth}{!}
{
    \centering
    \begin{tabular}{cccccccccccc}
    \toprule
    \multirow{2}[3]{*}{Model} & Dataset & \multicolumn{2}{c}{CIFAR10} & \multicolumn{2}{c}{GTSRB} & \multicolumn{2}{c}{CIFAR100} & \multicolumn{2}{c}{Tiny-ImageNet} & \multicolumn{2}{c}{ImageNet200} \\
\cmidrule{2-12}          & Metric & AVG   & STD   & AVG   & STD   & AVG   & STD   & AVG   & STD   & AVG   & STD \\
    \multirow{2}[1]{*}{CNNs} & AUROC & 0.918  & 0.000  & 0.930  & 0.012  & 0.884  & 0.055  & 0.970  & 0.000  & 0.949  & 0.019  \\
          & F1 score & 0.914  & 0.001  & 0.929  & 0.010  & 0.879  & 0.032  & 0.966  & 0.000  & 0.943  & 0.017  \\
    \midrule
    \multirow{2}[2]{*}{ViTs} & AUROC & 0.868  & 0.004  & 0.929  & 0.005  & 0.936  & 0.012  & 0.954  & 0.050  & 0.975  & 0.002  \\
          & F1 score & 0.857  & 0.004  & 0.931  & 0.006  & 0.928  & 0.016  & 0.946  & 0.033  & 0.969  & 0.003  \\
    \bottomrule
    \end{tabular}%
}
\setlength{\belowcaptionskip}{-0.2cm}
\caption{Performance of TeCo with a different maximum of severity. The average and standard deviation results suggest that TeCo has high effectiveness in different max severity $N$ and maintains stable performance among different $N$.}
\vspace{-0.3cm}
\label{max_table}%
\end{table}%

\begin{figure}[h]
\setlength{\belowcaptionskip}{-0.3cm}
\centering
    \begin{subfigure}{0.22\textwidth}
        \includegraphics[width=\textwidth]{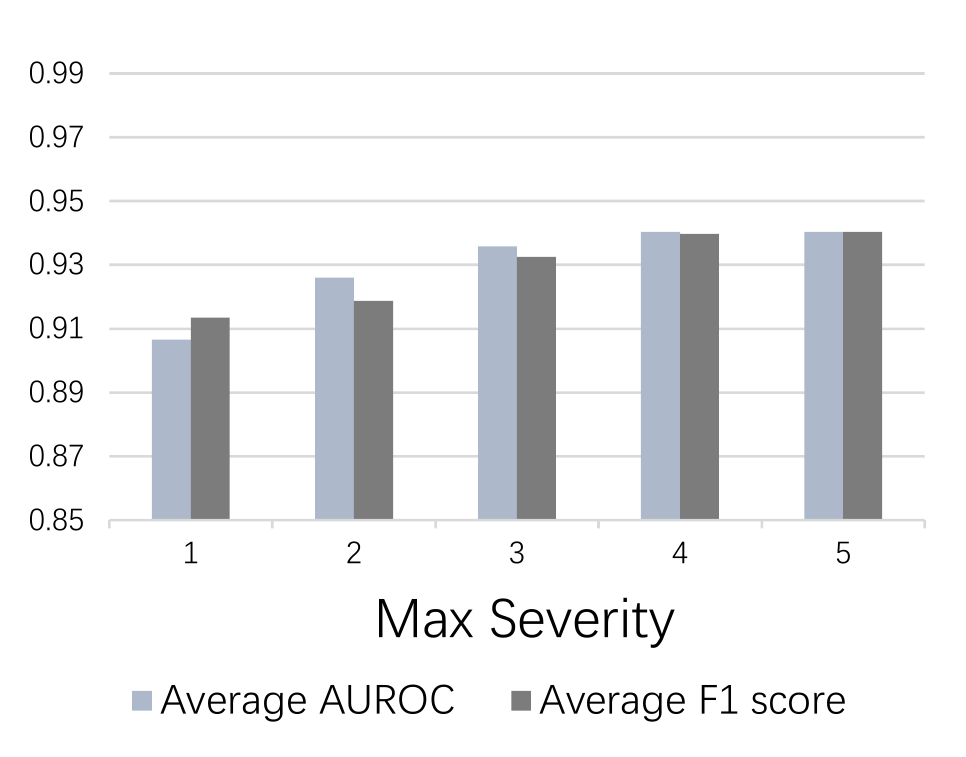} 
        \caption{PreActResNet18 / CIFAR10}
        \label{Fig.max_cifar10}
    \end{subfigure}
    \begin{subfigure}{0.22\textwidth}
        \includegraphics[width=\textwidth]{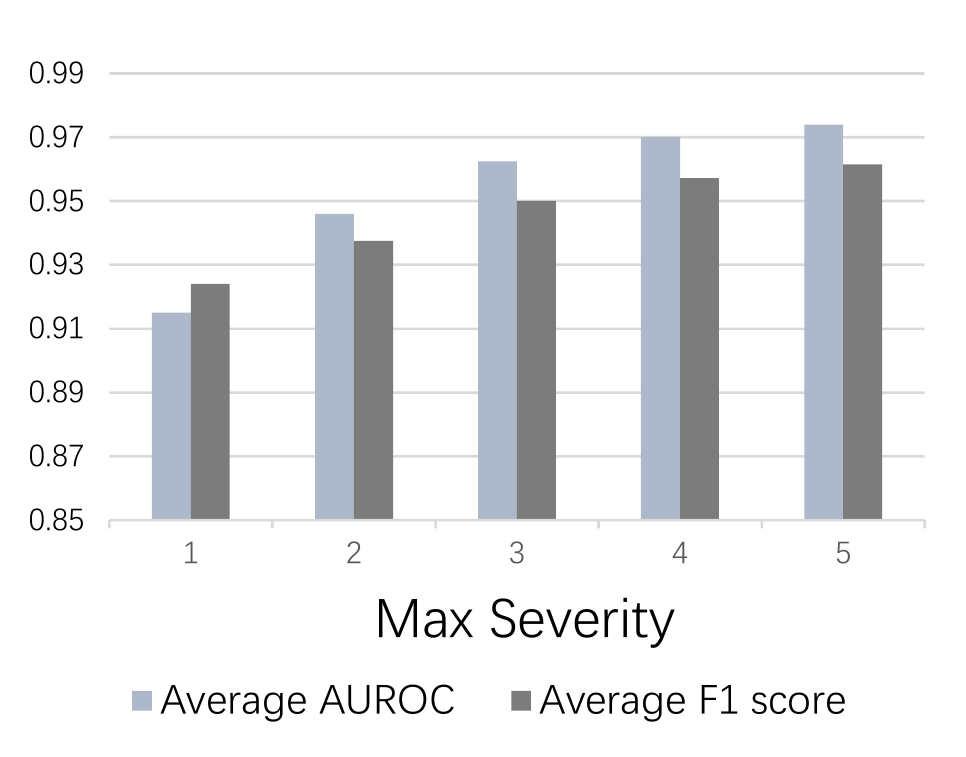} 
        \caption{MobileViT-xs / Tiny-Imagenet}
        \label{Fig.max_tiny}
    \end{subfigure}
\caption{Illustration of TeCo's performance with different max severity. Despite TeCo's performance grows with the increasing max severity $N$, TeCo still has good performance with low $N$.}
\vspace{-0.2cm}
\label{max_fig}
\end{figure}

\subsection{Thresholds}
Since TeCo maps the input image $x$ to a linearly separable space and defenders make judgments by a threshold $\gamma$, questions are how we can get this threshold and what is the influence of threshold for our method. Here, we evaluate TeCo by setting an empirical threshold directly, which does not break the ``no need for extra data" characteristic of TeCo. We use ACC as the evaluation metric, which is calculated by:
\begin{small}
\begin{equation}\label{eq:static_th}
\begin{aligned}
ACC = \frac{TP+TN}{TP+TN+FP+FN}.
\end{aligned}
\end{equation}
\end{small}

Tab.~\ref{empirical_table} shows the average performance of TeCo in different attacks, datasets, and backbones when an empirical threshold is given. The results suggest that even in the worst case where no data is available for defenders to estimate an appropriate threshold, by empirically setting threshold $=1$, TeCo can still get an average ACC $\approx0.79$, which still surpasses STRIP's performance in optimal thresholds shown in Tab.~\ref{main_table} and is competitive to FreqDetector. We defer a more detailed discussion about thresholds in other different settings to the supplementary. 

\begin{table}[h]
\centering
\vspace{-0.2cm}
  \resizebox{0.45\textwidth}{!}
  {
    \begin{tabular}{ccccccc}
    \toprule
          & CIFAR10 & GTSRB & CIFAR100 & Tiny-ImageNet & ImageNet200 & AVG \\
    \midrule
    CNNs  & 0.8521  & 0.9242  & 0.7735  & 0.6504  & 0.7613  & 0.7923  \\
    ViTs  & 0.8018  & 0.8366  & 0.7778  & 0.7569  & 0.7610  & 0.7868  \\
    \bottomrule
    \end{tabular}%
    }
    \vspace{-0.2cm}
\caption{The accuracy of TeCo in the settings where only one empirical threshold ($\gamma=1$) can be set for all attacks}
\label{empirical_table}%
\vspace{-0.5cm}
\end{table}%

\section{Analyses}\label{analysis}
\subsection{Constructing Adaptive Attacks against TeCo}
As formulated by Eq.(~\ref{eq:backdoor_attack}), the goal of backdoor attacks is to make models perform normally on clean data but give a specific prediction on trigger samples, the classic loss function for training such models can be defined as:
\begin{small}
\begin{equation}\label{eq:backdoor_loss}
\begin{aligned}
{\mathcal J_{bd}} = \sum^I_{i=1} CE(C_\theta(x_i), y_i) 
+ \sum^J_{j=1} CE(C_\theta(\hat{x_j}), y_t),
\end{aligned}
\end{equation}
\end{small}where $CE(\cdot)$ represents the cross entropy loss function. This backdoor loss function is widely used in backdoor attacks.
However, we speculate that this dual-target loss function leads backdoor-infected models to act anomalously on trigger samples in terms of corruption robustness. To reveal this point, we first introduce an adaptive loss to attack our method TeCo:
\begin{small}
\begin{equation}\label{eq:adaptive_loss}
\begin{aligned}
{\mathcal J_{ada}} = \sum^J_{j=1}\sum^K_{k=1}\sum^N_{n=1}&MSE(MSE(C_\theta(x_j), C_\theta(D^k_n(x_j))), \\
&MSE(C_\theta(\hat{x_j}), C_\theta(D^k_n(\hat{x_j})))),
\end{aligned}
\end{equation}
\end{small}where $x_j$ is the original version of the trigger sample $\hat{x_j}$. This adaptive loss aims to make models have the same corruption robustness on corrupted clean samples and corrupted trigger samples, which is aligned to the inference logic of TeCo.

\begin{figure}[t]
\setlength{\belowcaptionskip}{-0.2cm}
\centering
    \begin{subfigure}{0.15\textwidth}
        \includegraphics[width=\textwidth]{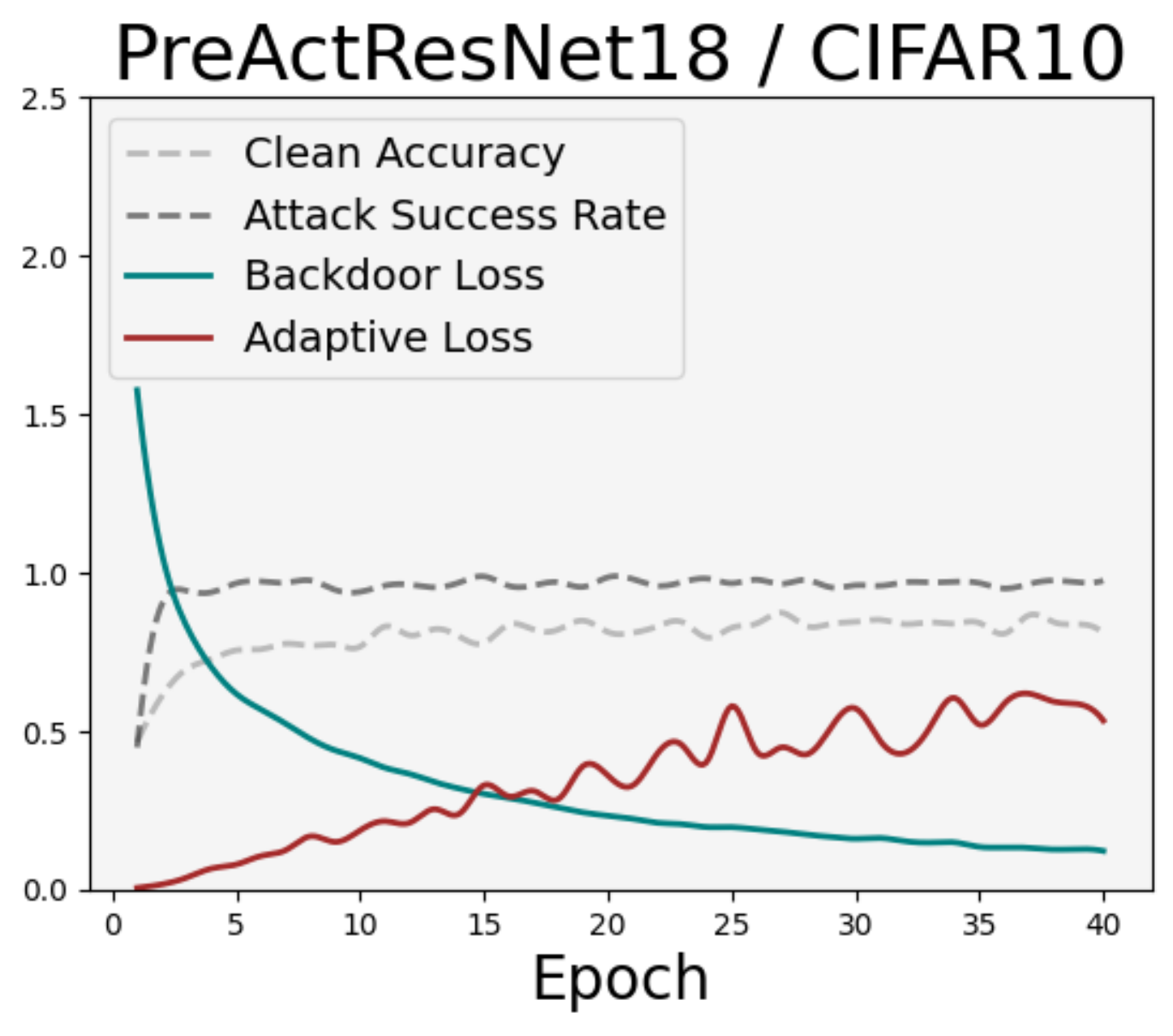} 
        \caption{Badnets}
        \label{Fig.badnet_adaptive}
    \end{subfigure}
    \begin{subfigure}{0.15\textwidth}
        \includegraphics[width=\textwidth]{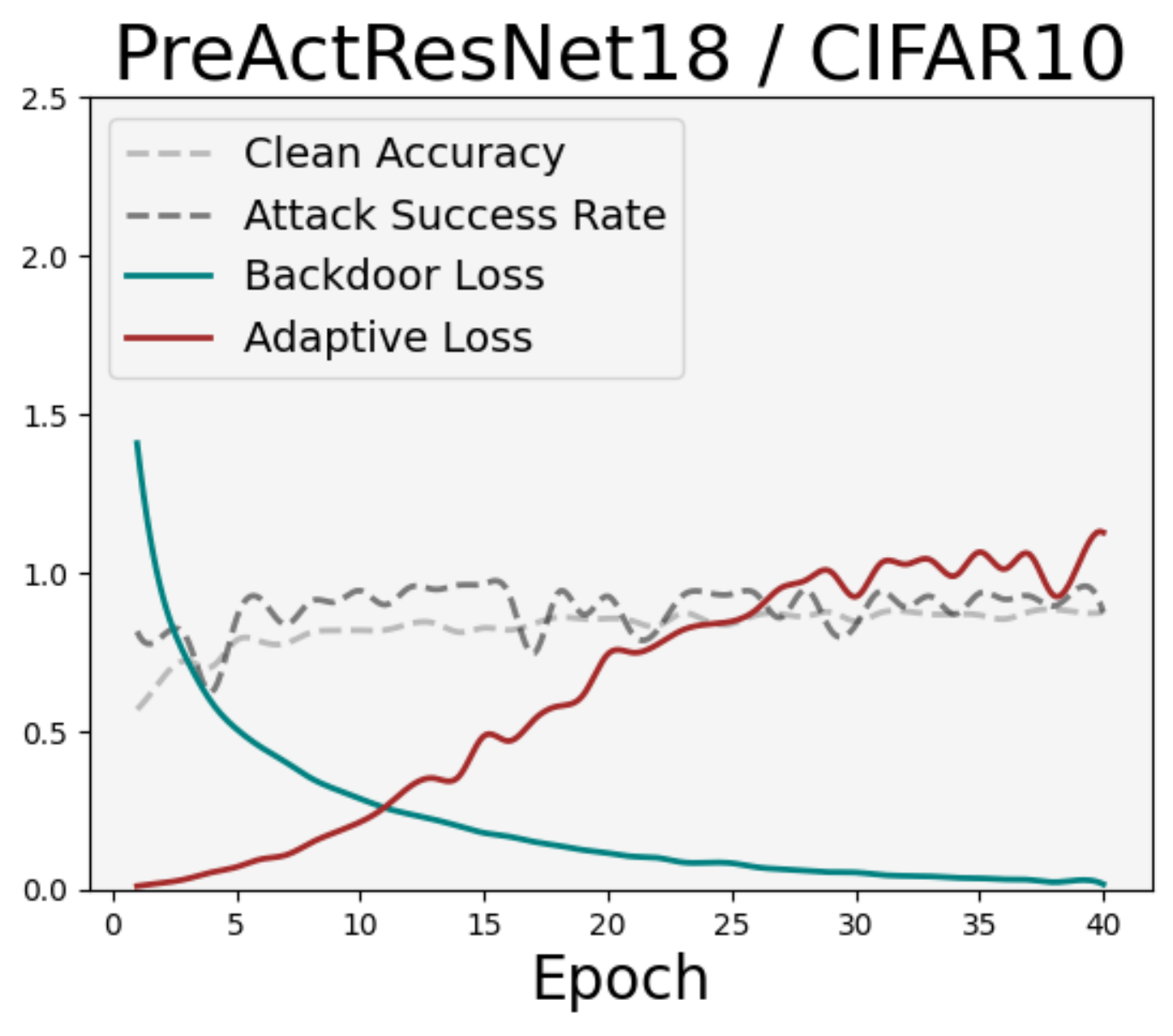} 
        \caption{LF}
        \label{Fig.lf_adaptive}
    \end{subfigure}
    \begin{subfigure}{0.15\textwidth}
        \includegraphics[width=\textwidth]{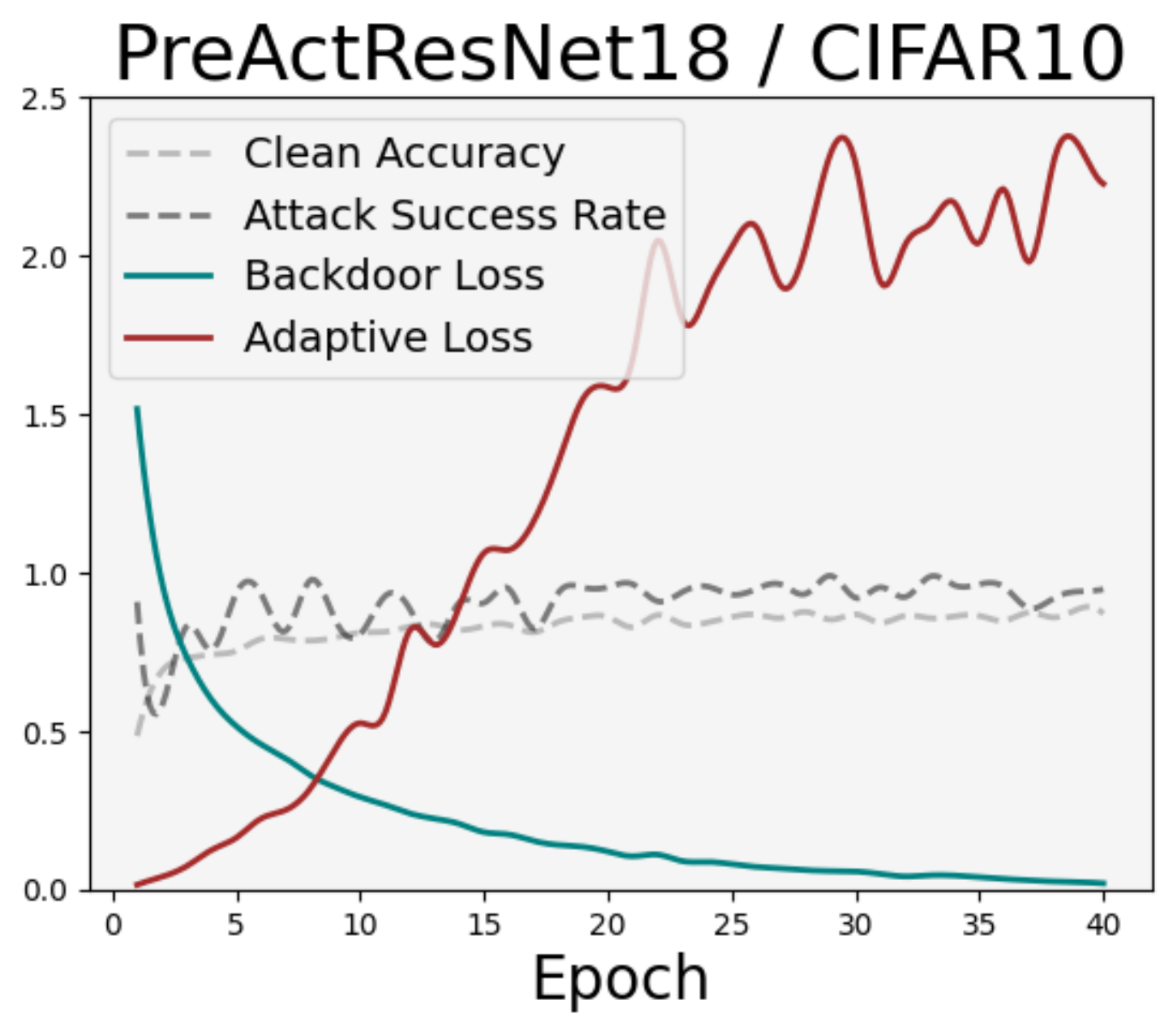} 
        \caption{SSBA}
        \label{Fig.ssba_adaptive}
    \end{subfigure}
\caption{Visualization of backdoor loss, adaptive loss, clean accuracy, and attack success rate in the training process. We note that with the drop of backdoor loss, the adaptive loss rises correspondingly, which indicates a negative correlation between them.}
\label{Fig.adaptive}
\end{figure}

\subsection{Results of Adaptive Attacks}
We convert Badnets, LF, and SSBA to the adaptive version by applying our adaptive loss in their training process, and investigate these adaptive attacks on PreActResNet18/CIFAR10. 
We first monitor the adaptive loss without derivative in the training phases. As illustrated in Fig.~\ref{Fig.adaptive}, the adaptive loss grows\footnote{We scale the adaptive loss down to fit the figure by multiplying $10^{-3}$.} when the backdoor loss decreases, which means the success on the dual-target loss function may drive the model to behave differently in terms of corruption robustness. A reasonable hypothesis is the model learns shortcuts~\cite{geirhos2020shortcut} for the backdoor trigger guided by the backdoor loss function, however, this trigger is not always robust in image space when facing different image corruptions. These results support TeCo's effectiveness by showing the negative correlation between backdoor loss and adaptive loss. And since the involved attacks contain different characteristics, such as partial (Badnets), global (LF), universal (Badnets, LF), sample-specific (SSBA), and invisible (SSBA), we suppose that this negative correlation is hard to be avoided by changing trigger types, which confirms the stability of TeCo on the other hand.

\begin{table}[t]
\centering
\resizebox{0.49\textwidth}{!}
{
    \begin{tabular}{ccccccccc}
    \toprule
    Weight→ & \multicolumn{2}{c}{$0$} & \multicolumn{2}{c}{$10^{-3}$} & \multicolumn{2}{c}{$10^{-4}$} & \multicolumn{2}{c}{$10^{-5}$} \\
    \midrule
    Attack↓ & AUROC & F1 score & AUROC & F1 score & AUROC & F1 score & AUROC & F1 score \\
    \midrule
    BadNets & 0.9112  & 0.9174  & 0.5763  & 0.5928  & 0.6571  & 0.6542  & 0.6745  & 0.6657  \\
    LF    & 0.9390  & 0.9367  & 0.8592  & 0.8483  & 0.9219  & 0.9154  & 0.8667  & 0.8858  \\
    SSBA  & 0.8683  & 0.8835  & 0.7125  & 0.7312  & 0.6477  & 0.7281  & 0.5909  & 0.6852  \\
    \bottomrule
    \\
    \toprule
    Weight→ & \multicolumn{2}{c}{$0$} & \multicolumn{2}{c}{$10^{-3}$} & \multicolumn{2}{c}{$10^{-4}$} & \multicolumn{2}{c}{$10^{-5}$} \\
    \midrule
    Attack↓ & C.ACC & ASR   & C.ACC & ASR   & C.ACC & ASR   & C.ACC & ASR \\
    \midrule
    BadNets & 0.9153  & 0.9502  & 0.5105  & 0.7386  & 0.7980  & 0.3720  & 0.8546  & 0.3001  \\
    LF    & 0.9286  & 0.9888  & 0.8022  & 0.9443  & 0.8864  & 0.9504  & 0.8962  & 0.9476  \\
    SSBA  & 0.9270  & 0.9719  & 0.7129  & 0.9176  & 0.8925  & 0.9162  & 0.8978  & 0.9170  \\
    \bottomrule
    \end{tabular}%
}
\vspace{-0.2cm}
\caption{Performance of TeCo against adaptive backdoor attacks}
\vspace{-0.4cm}
\label{tab:adaptive_roc}%
\end{table}%

\begin{figure}[t]
\centering
    \begin{subfigure}{0.2\textwidth}
        \includegraphics[width=\textwidth]{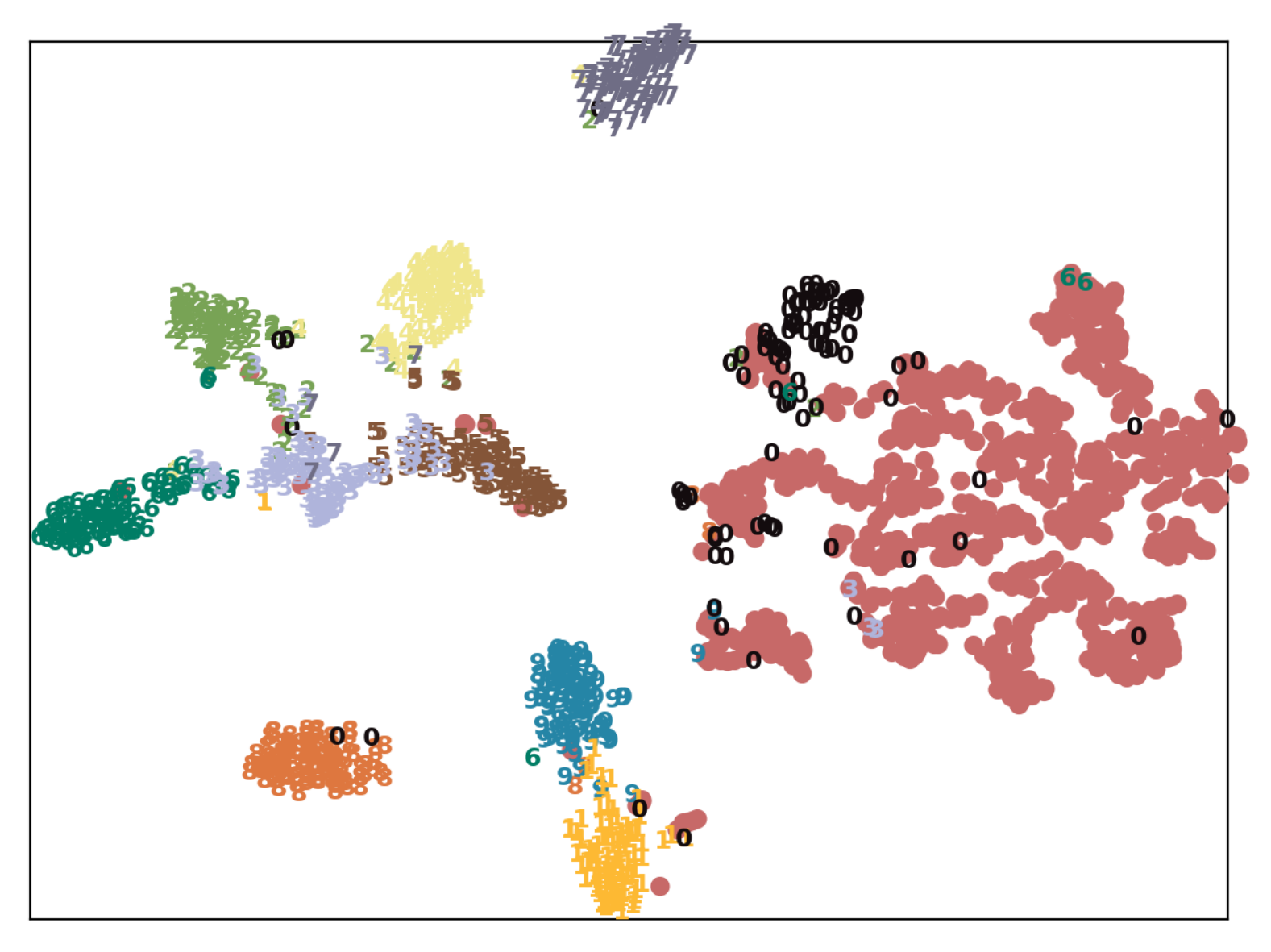} 
        \caption{Without adaptive loss}
        \label{feature_0}
    \end{subfigure}
    \begin{subfigure}{0.2\textwidth}
        \includegraphics[width=\textwidth]{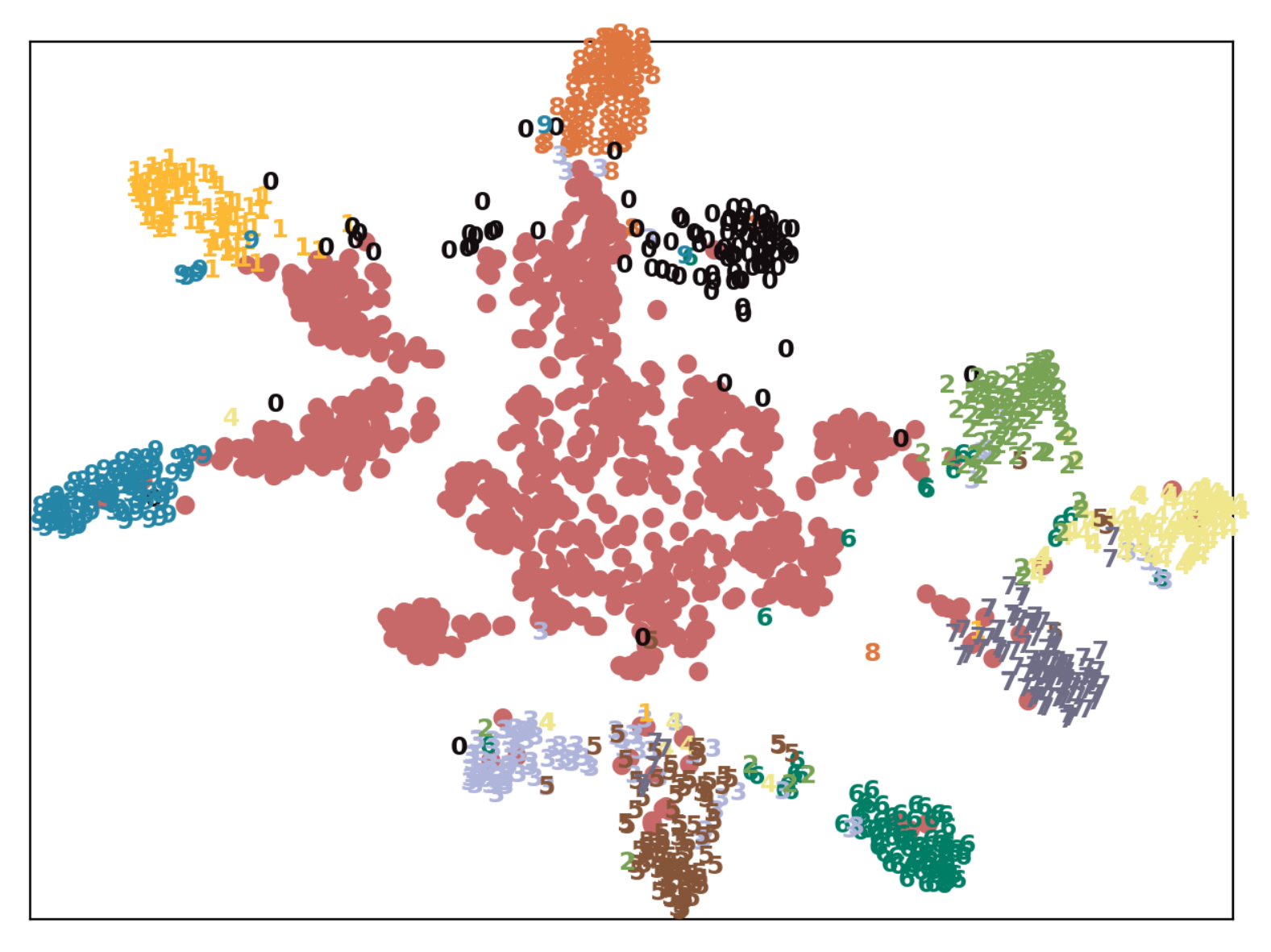} 
        \caption{$\alpha = 10^{-5}$}
        \label{feature_e5}
    \end{subfigure}
    \\
    \begin{subfigure}{0.2\textwidth}
        \includegraphics[width=\textwidth]{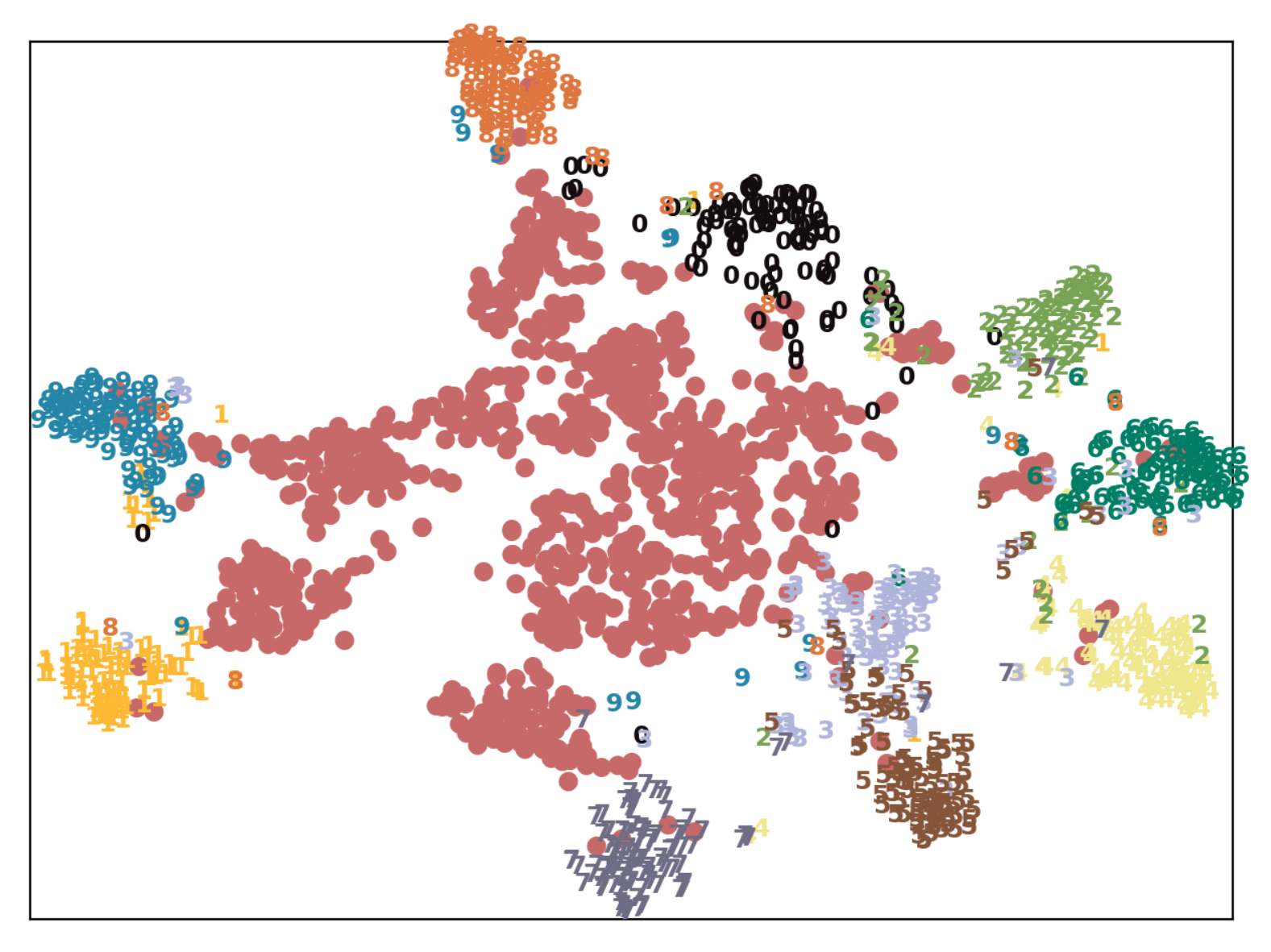} 
        \caption{$\alpha = 10^{-4}$}
        \label{feature_e4}
    \end{subfigure}
    \begin{subfigure}{0.2\textwidth}
        \includegraphics[width=\textwidth]{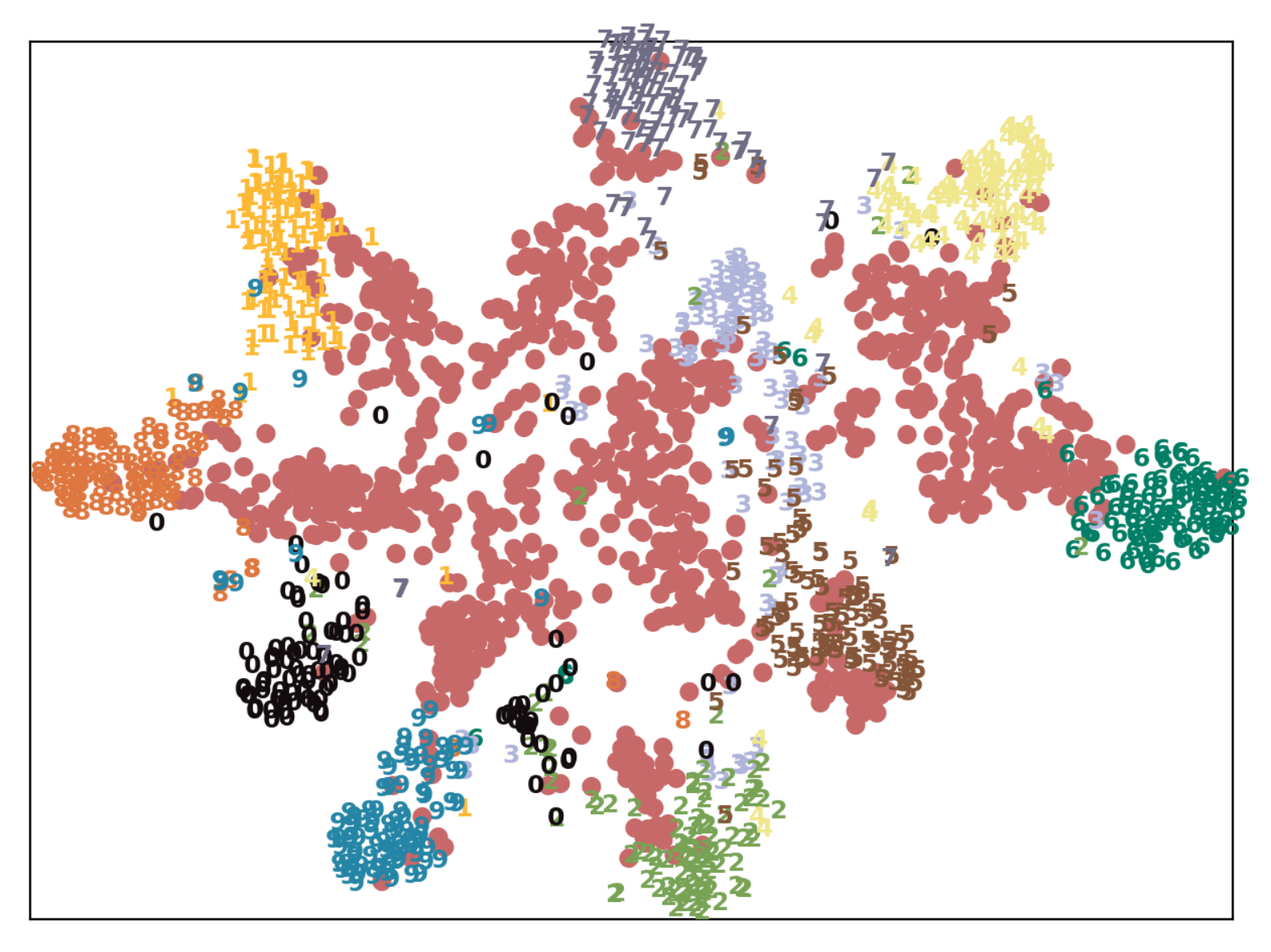} 
        \caption{$\alpha = 10^{-3}$}
        \label{feature_e3}
    \end{subfigure}
    \setlength{\belowcaptionskip}{-0.3cm}
    \vspace{-0.2cm}
    \caption{The red points represent the trigger samples, and the black points are clean samples from the target class. The points in other colors are clean samples from other classes.}
    \label{adaptive_features}
    \vspace{-0.2cm}
\end{figure}

Then we add the adaptive loss to the overall loss function: ${\mathcal J}={\mathcal J_{bd}}+\alpha{\mathcal J_{ada}}$, where $\alpha$ is the weight factor. Tab.~\ref{tab:adaptive_roc} shows the TeCo's effectiveness against adaptive attacks and the performance of adaptive attacks. The results indicate that the adaptive attacks can avoid TeCo to some degree, however they sacrifice attack performance once applying the adaptive loss. 

Since SSBA has the best performance in Tab.~\ref{tab:adaptive_roc}, we further visualize the clean and trigger samples in the latent space of the SSBA-infected model. As illustrated in Fig.~\ref{adaptive_features}, the adaptive loss pushes the trigger samples from the edge of latent space to the center, making them have a similar distance to different clean samples. Thus, a possible way to attack TeCo is to embed trigger samples in the middle of the latent space. However, this may be hard to achieve as we have shown in Tab.~\ref{adaptive_stable}, the proposed adaptive attack is not stable enough on different datasets.

\begin{table}[h]
\vspace{-0.2cm}
\setlength{\belowcaptionskip}{-0.3cm}
\centering
   \resizebox{0.47\textwidth}{!}
   {
    \begin{tabular}{ccccccc}
    \toprule
    Dataset & AUROC(↑) & F1 score(↑) & ACC(↑) & FAR(↓) & FRR(↓) & BDR(↑) \\
    \midrule
    GTSRB & 0.9101  & 0.8900  & 89.00  & 13.09  & 8.90  & 91.10  \\
    CIFAR100 & 0.9141  & 0.9137  & 91.37  & 5.76  & 11.54  & 88.46  \\
    \bottomrule
    \end{tabular}%
    }
\vspace{-0.3cm}
\caption{TeCo against adaptive SSBA attack ($10^{-5}$) on PreActResNet18}
\vspace{-0.2cm}
\label{adaptive_stable}%
\end{table}%

\section{Conclusions, Limitations, and Future Work}\label{Limitation}
\vspace{-0.2cm}
In this paper, we propose TeCo, the first test-time trigger sample detection method that only needs the hard-label outputs of the victim models without requiring extra data or assumptions. Extensive experiments support that TeCo has outstanding effectiveness and stability against different backdoor attacks. However, a limitation of TeCo is that using multiple image corruptions will increase the computational cost. Therefore, designing an effective and efficient single corruption function will be our future work.

~\\
\vspace{-0.05cm}

\noindent\textbf{Acknowledgments.} Shengshan's work is supported in part by the National Natural Science Foundation of China (Grant No.U20A20177) and Hubei Province Key R\&D Technology Special Innovation Project under Grant No.2021BAA032. Shengshan Hu is the corresponding author.

{\small
\bibliographystyle{ieee_fullname}
\bibliography{egbib}
}

\clearpage
\renewcommand\thesection{\Alph{section}}
\setcounter{section}{0}
\section{Implementation Details}\label{details}
\begin{table*}[htbp]
\setlength{\belowcaptionskip}{-0.2cm}
  \resizebox{1.0\textwidth}{!}
  {
  \centering
    \begin{tabular}{lccccccccccccccc}
    \toprule
    \multicolumn{1}{c}{\multirow{2}[2]{*}{Dataset}} & Attack→ & \multicolumn{2}{c}{Badnets} & \multicolumn{2}{c}{Blended} & \multicolumn{2}{c}{LF} & \multicolumn{2}{c}{Input-aware} & \multicolumn{2}{c}{Wanet} & \multicolumn{2}{c}{LIRA} & \multicolumn{2}{c}{SSBA} \\
          & Backbone↓ & ACC   & ASR   & ACC   & ASR   & ACC   & ASR   & ACC   & ASR   & ACC   & ASR   & ACC   & ASR   & ACC   & ASR \\
    \midrule
    \multicolumn{1}{c}{\multirow{2}[2]{*}{CIFAR10}} & PreActResNet18 & 91.53  & 95.02  & 93.09  & 99.71  & 92.86  & 98.88  & 90.33  & 94.50  & 90.37  & 91.23  & 89.94  & 100.00  & 92.70  & 97.19  \\
          & MobileViT-xs & 90.62  & 95.71  & 91.14  & 99.50  & 90.67  & 96.37  & 87.84  & 96.67  & 88.94  & 90.78  & 83.89  & 100.00  & 90.29  & 95.28  \\
    \midrule
    \multicolumn{1}{c}{\multirow{2}[2]{*}{GTSRB}} & PreActResNet18 & 97.74  & 93.35  & 98.20  & 99.98  & 97.25  & 99.86  & 97.36  & 96.39  & 97.74  & 92.94  & 96.37  & 100.00  & 98.23  & 99.53  \\
          & MobileViT-xs & 97.52  & 94.48  & 97.49  & 99.98  & 97.82  & 98.35  & 96.53  & 97.21  & 95.44  & 94.77  & 93.97  & 100.00  & 97.65  & 98.72  \\
    \midrule
    \multicolumn{1}{c}{\multirow{2}[2]{*}{CIFAR100}} & PreActResNet18 & 67.38  & 88.09  & 69.63  & 99.45  & 68.96  & 94.71  & 64.48  & 88.46  & 64.43  & 93.41  & 66.42  & 100.00  & 68.81  & 97.54  \\
          & MobileViT-xs & 59.62  & 89.39  & 61.95  & 99.52  & 61.36  & 95.45  & 55.63  & 92.38  & 59.24  & 75.81  & 52.98  & 100.00  & 60.80  & 96.87  \\
    \midrule
    \multicolumn{1}{c}{\multirow{2}[2]{*}{Tiny-ImageNet}} & PreActResNet18 & 56.11  & 99.97  & 56.40  & 99.59  & 55.74  & 98.64  & 57.09  & 99.08  & 57.29  & 99.51  & 54.57  & 99.96  & 55.32  & 97.73  \\
          & MobileViT-xs & 47.61  & 99.99  & 48.08  & 99.90  & 48.41  & 97.18  & 55.91  & 99.67  & 55.38  & 99.18  & 51.00  & 99.95  & 48.24  & 97.27  \\
    \midrule
    \multicolumn{1}{c}{\multirow{2}[2]{*}{ImageNet200}} & WideResNet101-2 & 71.06  & 99.76  & 71.75  & 99.28  & -     & -     & 75.65  & 82.04  & 94.44  & 90.36  & 77.39  & 100.00  & 90.51  & 94.14  \\
          & SwinT-Base & 74.48  & 99.94  & 78.89  & 100.00  & -     & -     & 84.92  & 99.91  & 77.04  & 94.83  & 82.88  & 100.00  & 97.50  & 86.22  \\
    \midrule
    GTSRB (all2all) & PreActResNet18 & 97.84  & 91.88  & 98.54  & 95.72  & 98.16  & 96.56  & 97.25  & 85.78  & 98.88  & 98.82  & 96.64  & 96.59  & 97.88  & 95.43  \\
    \bottomrule
    \end{tabular}%
    }
\caption{The effectiveness of backdoor attacks on different backbones and datasets. We use these backdoor-infected models to further evaluate our method.} 
\label{attacks}%
\vspace{-0.3cm}
\end{table*}%

\begin{table}[htbp]
\resizebox{0.47\textwidth}{!}
{
    \centering
    \begin{tabular}{cccccccccccc}
    \toprule
    Attack & Metric & 1     & 2     & 3     & 4     & 5     & 6     & 7     & 8     & 9     & 10 \\
    \midrule
    \multirow{2}[2]{*}{Badnets} & ASR   & 93.35  & 95.52  & 95.76  & 94.93  & 95.16  & 94.59  & 94.92  & 96.12  & 94.54  & 95.57  \\
          & ACC   & 97.74  & 97.53  & 97.86  & 97.54  & 97.77  & 97.42  & 97.77  & 97.78  & 97.66  & 97.21  \\
    \midrule
    \multirow{2}[2]{*}{Input-aware} & ASR   & 92.94  & 92.73  & 90.84  & 95.07  & 90.56  & 96.00  & 97.01  & 93.46  & 92.45  & 94.40  \\
          & ACC   & 97.74  & 98.69  & 98.60  & 98.39  & 98.71  & 98.31  & 97.76  & 97.94  & 98.16  & 97.19  \\
    \midrule
    \multirow{2}[2]{*}{Wanet} & ASR   & 96.39  & 95.65  & 92.56  & 89.37  & 90.89  & 93.41  & 99.33  & 96.62  & 97.27  & 98.17  \\
          & ACC   & 97.36  & 97.36  & 97.48  & 98.65  & 97.81  & 98.57  & 98.13  & 97.47  & 97.13  & 96.94  \\
    \bottomrule
    \end{tabular}%
}
\caption{The effectiveness of backdoor attacks on different target labels}
\label{targets}%
\end{table}%

\begin{table}[htbp]
\resizebox{0.47\textwidth}{!}
{
    \centering
    \begin{tabular}{cccccc}
    \toprule
    \multirow{2}[4]{*}{Dataset} & \multirow{2}[4]{*}{\#Classes} & \multirow{2}[4]{*}{Image Size} & \multirow{2}[4]{*}{Training Data} & \multicolumn{2}{c}{Test Data} \\
\cmidrule{5-6}          &       &       &       & Clean Images & Trigger Samples \\
    \midrule
    CIFAR10 & 10    & 3×32×32 & 50000 & 10000 & 9000 \\
    GTSRB & 43    & 3×32×32 & 39209 & 12630 & 12570 \\
    CIFAR100 & 100   & 3×32×32 & 50000 & 10000 & 9900 \\
    Tiny-ImageNet & 200   & 3×64×64 & 100000 & 10000 & 9950 \\
    ImageNet200 & 200   & 3×224×224 & 100000 & 10000 & 9950 \\
    \bottomrule
    \end{tabular}%
}
\setlength{\belowcaptionskip}{-0.4cm}
\caption{Datasets for evaluations}
\label{datasets}%
\end{table}%

\subsection{Baselines}
Tab.~\ref{attacks} and Tab.~\ref{targets} show the effectiveness of different attacks on different backbones and datasets, indicating that all the attacks are valid.

\textbf{STRIP~\cite{gao2019strip}. }We re-implement STRIP following the official codes~\footnote{\url{https://github.com/garrisongys/STRIP}} and a reference~\footnote{\url{https://github.com/wanlunsec/Beatrix/tree/master/defenses/STRIP}}. For every input image, we use $100$ clean images from test data for superimposing.

\textbf{FreqDetector~\cite{zeng2021rethinking}. }We re-implement FreqDetector following the official codes~\footnote{\url{https://github.com/YiZeng623/frequency-backdoor/tree/main/Sec4_Frequency_Detection}}. We choose PreActResNet18 as the backbone of FreqDetector, and let all clean training images (for example, $50000$ images in CIFAR10) serve as the training data of FreqDetector. Following the paper and official codes, we choose a random white block, random colored block, Gaussian noise, random shadow, and random blend as data augmentations.

\begin{table*}[htbp]
\setlength{\belowcaptionskip}{-0.2cm}
  \resizebox{1.0\textwidth}{!}
  {
  \centering
    \begin{threeparttable}
        \begin{tabular}{ccccccccccccccccccccccccccc}
    \toprule
    \multirow{2}[2]{*}{Dataset} & \multirow{2}[2]{*}{Model} & Attack→ & \multicolumn{3}{c}{Badnets} & \multicolumn{3}{c}{Blended} & \multicolumn{3}{c}{LF} & \multicolumn{3}{c}{Input-Aware} & \multicolumn{3}{c}{Wanet} & \multicolumn{3}{c}{LIRA} & \multicolumn{3}{c}{SSBA} & \multicolumn{3}{c}{AVG} \\
          &       & Detection↓ & FAR   & FRR   & BDR   & FAR   & FRR   & BDR   & FAR   & FRR   & BDR   & FAR   & FRR   & BDR   & FAR   & FRR   & BDR   & FAR   & FRR   & BDR   & FAR   & FRR   & BDR   & FAR   & FRR   & BDR \\
    \midrule
    \multirow{6}[4]{*}{CIFAR10} & \multirow{3}[2]{*}{PreActResNet18} & STRIP & 0.37  & 0.15  & 0.85  & 0.38  & 0.26  & 0.74  & 0.08  & 0.05  & 0.95  & 1.00  & 0.00  & 1.00  & 1.00  & 0.00  & 1.00  & 0.71  & 0.01  & 0.99  & 1.00  & 0.00  & 1.00  & 0.65  & 0.07  & 0.93  \\
          &       & FreqDetector & 0.02  & 0.08  & 0.92  & 0.07  & 0.12  & 0.88  & 0.10  & 0.29  & 0.71  & 0.01  & 0.01  & 0.99  & 0.38  & 0.52  & 0.48  & 0.10  & 0.23  & 0.77  & 0.10  & 0.26  & 0.74  & 0.11  & 0.22  & 0.78  \\
          &       & \cellcolor[rgb]{ .906,  .902,  .902}Ours & \cellcolor[rgb]{ .906,  .902,  .902}0.11  & \cellcolor[rgb]{ .906,  .902,  .902}0.05  & \cellcolor[rgb]{ .906,  .902,  .902}0.95  & \cellcolor[rgb]{ .906,  .902,  .902}0.10  & \cellcolor[rgb]{ .906,  .902,  .902}0.00  & \cellcolor[rgb]{ .906,  .902,  .902}1.00  & \cellcolor[rgb]{ .906,  .902,  .902}0.11  & \cellcolor[rgb]{ .906,  .902,  .902}0.01  & \cellcolor[rgb]{ .906,  .902,  .902}0.99  & \cellcolor[rgb]{ .906,  .902,  .902}0.10  & \cellcolor[rgb]{ .906,  .902,  .902}0.06  & \cellcolor[rgb]{ .906,  .902,  .902}0.95  & \cellcolor[rgb]{ .906,  .902,  .902}0.10  & \cellcolor[rgb]{ .906,  .902,  .902}0.09  & \cellcolor[rgb]{ .906,  .902,  .902}0.91  & \cellcolor[rgb]{ .906,  .902,  .902}0.12  & \cellcolor[rgb]{ .906,  .902,  .902}0.01  & \cellcolor[rgb]{ .906,  .902,  .902}0.99  & \cellcolor[rgb]{ .906,  .902,  .902}0.20  & \cellcolor[rgb]{ .906,  .902,  .902}0.03  & \cellcolor[rgb]{ .906,  .902,  .902}0.97  & \cellcolor[rgb]{ .906,  .902,  .902}0.12  & \cellcolor[rgb]{ .906,  .902,  .902}0.04  & \cellcolor[rgb]{ .906,  .902,  .902}0.97  \\
\cmidrule{2-27}          & \multirow{3}[2]{*}{MobileViT-xs} & STRIP & 0.44  & 0.16  & 0.84  & 0.76  & 0.17  & 0.83  & 0.14  & 0.14  & 0.86  & 1.00  & 0.00  & 1.00  & 1.00  & 0.00  & 1.00  & 0.86  & 0.01  & 1.00  & 1.00  & 0.00  & 1.00  & 0.74  & 0.07  & 0.93  \\
          &       & FreqDetector & 0.02  & 0.08  & 0.92  & 0.07  & 0.12  & 0.88  & 0.13  & 0.35  & 0.65  & 0.03  & 0.03  & 0.97  & 0.00  & 1.00  & 0.00  & 0.03  & 0.10  & 0.90  & 0.10  & 0.26  & 0.74  & 0.05  & 0.28  & 0.72  \\
          &       & \cellcolor[rgb]{ .906,  .902,  .902}Ours & \cellcolor[rgb]{ .906,  .902,  .902}0.44  & \cellcolor[rgb]{ .906,  .902,  .902}0.10  & \cellcolor[rgb]{ .906,  .902,  .902}0.90  & \cellcolor[rgb]{ .906,  .902,  .902}0.14  & \cellcolor[rgb]{ .906,  .902,  .902}0.01  & \cellcolor[rgb]{ .906,  .902,  .902}0.99  & \cellcolor[rgb]{ .906,  .902,  .902}0.14  & \cellcolor[rgb]{ .906,  .902,  .902}0.04  & \cellcolor[rgb]{ .906,  .902,  .902}0.96  & \cellcolor[rgb]{ .906,  .902,  .902}0.22  & \cellcolor[rgb]{ .906,  .902,  .902}0.21  & \cellcolor[rgb]{ .906,  .902,  .902}0.79  & \cellcolor[rgb]{ .906,  .902,  .902}0.10  & \cellcolor[rgb]{ .906,  .902,  .902}0.09  & \cellcolor[rgb]{ .906,  .902,  .902}0.91  & \cellcolor[rgb]{ .906,  .902,  .902}0.07  & \cellcolor[rgb]{ .906,  .902,  .902}0.07  & \cellcolor[rgb]{ .906,  .902,  .902}0.93  & \cellcolor[rgb]{ .906,  .902,  .902}0.12  & \cellcolor[rgb]{ .906,  .902,  .902}0.06  & \cellcolor[rgb]{ .906,  .902,  .902}0.95  & \cellcolor[rgb]{ .906,  .902,  .902}0.17  & \cellcolor[rgb]{ .906,  .902,  .902}0.08  & \cellcolor[rgb]{ .906,  .902,  .902}0.92  \\
    \midrule
    \multirow{6}[4]{*}{GTSRB} & \multirow{3}[2]{*}{PreActResNet18} & STRIP & 0.24  & 0.08  & 0.92  & 0.22  & 0.08  & 0.92  & 0.03  & 0.01  & 0.99  & 1.00  & 0.00  & 1.00  & 1.00  & 0.00  & 1.00  & 0.40  & 0.02  & 0.98  & 0.41  & 0.34  & 0.66  & 0.47  & 0.08  & 0.92  \\
          &       & FreqDetector & 0.02  & 0.10  & 0.90  & 0.04  & 0.04  & 0.96  & 0.12  & 0.08  & 0.92  & 0.12  & 0.18  & 0.82  & 0.88  & 0.11  & 0.89  & 0.48  & 0.40  & 0.60  & 0.14  & 0.77  & 0.23  & 0.26  & 0.24  & 0.76  \\
          &       & \cellcolor[rgb]{ .906,  .902,  .902}Ours & \cellcolor[rgb]{ .906,  .902,  .902}0.18  & \cellcolor[rgb]{ .906,  .902,  .902}0.15  & \cellcolor[rgb]{ .906,  .902,  .902}0.85  & \cellcolor[rgb]{ .906,  .902,  .902}0.12  & \cellcolor[rgb]{ .906,  .902,  .902}0.05  & \cellcolor[rgb]{ .906,  .902,  .902}0.95  & \cellcolor[rgb]{ .906,  .902,  .902}0.07  & \cellcolor[rgb]{ .906,  .902,  .902}0.01  & \cellcolor[rgb]{ .906,  .902,  .902}0.99  & \cellcolor[rgb]{ .906,  .902,  .902}0.05  & \cellcolor[rgb]{ .906,  .902,  .902}0.04  & \cellcolor[rgb]{ .906,  .902,  .902}0.96  & \cellcolor[rgb]{ .906,  .902,  .902}0.01  & \cellcolor[rgb]{ .906,  .902,  .902}0.07  & \cellcolor[rgb]{ .906,  .902,  .902}0.93  & \cellcolor[rgb]{ .906,  .902,  .902}0.03  & \cellcolor[rgb]{ .906,  .902,  .902}0.00  & \cellcolor[rgb]{ .906,  .902,  .902}1.00  & \cellcolor[rgb]{ .906,  .902,  .902}0.06  & \cellcolor[rgb]{ .906,  .902,  .902}0.01  & \cellcolor[rgb]{ .906,  .902,  .902}0.99  & \cellcolor[rgb]{ .906,  .902,  .902}0.07  & \cellcolor[rgb]{ .906,  .902,  .902}0.05  & \cellcolor[rgb]{ .906,  .902,  .902}0.95  \\
\cmidrule{2-27}          & \multirow{3}[2]{*}{MobileViT-xs} & STRIP & 0.02  & 0.11  & 0.89  & 0.24  & 0.04  & 0.96  & 0.11  & 0.02  & 0.98  & 1.00  & 0.00  & 1.00  & 1.00  & 0.00  & 1.00  & 0.62  & 0.01  & 0.99  & 0.61  & 0.29  & 0.71  & 0.51  & 0.07  & 0.93  \\
          &       & FreqDetector & 0.02  & 0.10  & 0.90  & 0.04  & 0.04  & 0.96  & 0.18  & 0.14  & 0.86  & 0.00  & 0.00  & 1.00  & 0.85  & 0.13  & 0.87  & 0.27  & 0.16  & 0.84  & 0.14  & 0.77  & 0.23  & 0.21  & 0.19  & 0.81  \\
          &       & \cellcolor[rgb]{ .906,  .902,  .902}Ours & \cellcolor[rgb]{ .906,  .902,  .902}0.15  & \cellcolor[rgb]{ .906,  .902,  .902}0.04  & \cellcolor[rgb]{ .906,  .902,  .902}0.96  & \cellcolor[rgb]{ .906,  .902,  .902}0.13  & \cellcolor[rgb]{ .906,  .902,  .902}0.00  & \cellcolor[rgb]{ .906,  .902,  .902}1.00  & \cellcolor[rgb]{ .906,  .902,  .902}0.01  & \cellcolor[rgb]{ .906,  .902,  .902}0.02  & \cellcolor[rgb]{ .906,  .902,  .902}0.98  & \cellcolor[rgb]{ .906,  .902,  .902}0.18  & \cellcolor[rgb]{ .906,  .902,  .902}0.06  & \cellcolor[rgb]{ .906,  .902,  .902}0.94  & \cellcolor[rgb]{ .906,  .902,  .902}0.03  & \cellcolor[rgb]{ .906,  .902,  .902}0.05  & \cellcolor[rgb]{ .906,  .902,  .902}0.95  & \cellcolor[rgb]{ .906,  .902,  .902}0.05  & \cellcolor[rgb]{ .906,  .902,  .902}0.05  & \cellcolor[rgb]{ .906,  .902,  .902}0.95  & \cellcolor[rgb]{ .906,  .902,  .902}0.07  & \cellcolor[rgb]{ .906,  .902,  .902}0.01  & \cellcolor[rgb]{ .906,  .902,  .902}0.99  & \cellcolor[rgb]{ .906,  .902,  .902}0.09  & \cellcolor[rgb]{ .906,  .902,  .902}0.03  & \cellcolor[rgb]{ .906,  .902,  .902}0.97  \\
    \midrule
    \multirow{6}[4]{*}{CIFAR100} & \multirow{3}[2]{*}{PreActResNet18} & STRIP & 0.25  & 0.12  & 0.88  & 0.36  & 0.20  & 0.80  & 0.11  & 0.10  & 0.90  & 1.00  & 0.00  & 1.00  & 1.00  & 0.00  & 1.00  & 0.76  & 0.06  & 0.94  & 0.41  & 0.29  & 0.71  & 0.56  & 0.11  & 0.89  \\
          &       & FreqDetector & 0.02  & 0.13  & 0.87  & 0.09  & 0.11  & 0.89  & 0.08  & 0.35  & 0.65  & 0.02  & 0.03  & 0.97  & 0.00  & 1.00  & 0.00  & 0.07  & 0.14  & 0.86  & 0.12  & 0.26  & 0.74  & 0.06  & 0.29  & 0.71  \\
          &       & \cellcolor[rgb]{ .906,  .902,  .902}Ours & \cellcolor[rgb]{ .906,  .902,  .902}0.04  & \cellcolor[rgb]{ .906,  .902,  .902}0.12  & \cellcolor[rgb]{ .906,  .902,  .902}0.88  & \cellcolor[rgb]{ .906,  .902,  .902}0.06  & \cellcolor[rgb]{ .906,  .902,  .902}0.05  & \cellcolor[rgb]{ .906,  .902,  .902}0.95  & \cellcolor[rgb]{ .906,  .902,  .902}0.07  & \cellcolor[rgb]{ .906,  .902,  .902}0.25  & \cellcolor[rgb]{ .906,  .902,  .902}0.75  & \cellcolor[rgb]{ .906,  .902,  .902}0.08  & \cellcolor[rgb]{ .906,  .902,  .902}0.17  & \cellcolor[rgb]{ .906,  .902,  .902}0.83  & \cellcolor[rgb]{ .906,  .902,  .902}0.02  & \cellcolor[rgb]{ .906,  .902,  .902}0.06  & \cellcolor[rgb]{ .906,  .902,  .902}0.94  & \cellcolor[rgb]{ .906,  .902,  .902}0.21  & \cellcolor[rgb]{ .906,  .902,  .902}0.14  & \cellcolor[rgb]{ .906,  .902,  .902}0.86  & \cellcolor[rgb]{ .906,  .902,  .902}0.04  & \cellcolor[rgb]{ .906,  .902,  .902}0.02  & \cellcolor[rgb]{ .906,  .902,  .902}0.98  & \cellcolor[rgb]{ .906,  .902,  .902}0.07  & \cellcolor[rgb]{ .906,  .902,  .902}0.12  & \cellcolor[rgb]{ .906,  .902,  .902}0.88  \\
\cmidrule{2-27}          & \multirow{3}[2]{*}{MobileViT-xs} & STRIP & 0.29  & 0.11  & 0.89  & 0.31  & 0.20  & 0.80  & 0.09  & 0.14  & 0.86  & 0.88  & 0.09  & 0.91  & 1.00  & 0.00  & 1.00  & 0.60  & 0.13  & 0.87  & 0.24  & 0.26  & 0.74  & 0.49  & 0.13  & 0.87  \\
          &       & FreqDetector & 0.02  & 0.13  & 0.87  & 0.09  & 0.11  & 0.89  & 0.09  & 0.23  & 0.77  & 0.01  & 0.01  & 0.99  & 0.00  & 1.00  & 0.00  & 0.08  & 0.16  & 0.84  & 0.12  & 0.26  & 0.74  & 0.06  & 0.27  & 0.73  \\
          &       & \cellcolor[rgb]{ .906,  .902,  .902}Ours & \cellcolor[rgb]{ .906,  .902,  .902}0.06  & \cellcolor[rgb]{ .906,  .902,  .902}0.12  & \cellcolor[rgb]{ .906,  .902,  .902}0.88  & \cellcolor[rgb]{ .906,  .902,  .902}0.07  & \cellcolor[rgb]{ .906,  .902,  .902}0.02  & \cellcolor[rgb]{ .906,  .902,  .902}0.98  & \cellcolor[rgb]{ .906,  .902,  .902}0.02  & \cellcolor[rgb]{ .906,  .902,  .902}0.05  & \cellcolor[rgb]{ .906,  .902,  .902}0.95  & \cellcolor[rgb]{ .906,  .902,  .902}0.07  & \cellcolor[rgb]{ .906,  .902,  .902}0.06  & \cellcolor[rgb]{ .906,  .902,  .902}0.94  & \cellcolor[rgb]{ .906,  .902,  .902}0.08  & \cellcolor[rgb]{ .906,  .902,  .902}0.16  & \cellcolor[rgb]{ .906,  .902,  .902}0.84  & \cellcolor[rgb]{ .906,  .902,  .902}0.04  & \cellcolor[rgb]{ .906,  .902,  .902}0.03  & \cellcolor[rgb]{ .906,  .902,  .902}0.97  & \cellcolor[rgb]{ .906,  .902,  .902}0.05  & \cellcolor[rgb]{ .906,  .902,  .902}0.04  & \cellcolor[rgb]{ .906,  .902,  .902}0.96  & \cellcolor[rgb]{ .906,  .902,  .902}0.06  & \cellcolor[rgb]{ .906,  .902,  .902}0.07  & \cellcolor[rgb]{ .906,  .902,  .902}0.93  \\
    \midrule
    \multirow{6}[4]{*}{Tiny-ImageNet} & \multirow{3}[2]{*}{PreActResNet18} & STRIP & 0.14  & 0.29  & 0.71  & 0.14  & 0.08  & 0.92  & 0.03  & 0.02  & 0.98  & 0.98  & 0.02  & 0.98  & 0.31  & 0.41  & 0.59  & 0.91  & 0.00  & 1.00  & 0.36  & 0.19  & 0.81  & 0.41  & 0.14  & 0.86  \\
          &       & FreqDetector & 0.25  & 0.44  & 0.56  & 0.01  & 0.01  & 0.99  & 0.19  & 0.16  & 0.84  & 0.00  & 0.00  & 1.00  & 0.49  & 0.28  & 0.72  & 0.03  & 0.15  & 0.85  & 0.03  & 0.05  & 0.95  & 0.15  & 0.15  & 0.85  \\
          &       & \cellcolor[rgb]{ .906,  .902,  .902}Ours & \cellcolor[rgb]{ .906,  .902,  .902}0.03  & \cellcolor[rgb]{ .906,  .902,  .902}0.00  & \cellcolor[rgb]{ .906,  .902,  .902}1.00  & \cellcolor[rgb]{ .906,  .902,  .902}0.04  & \cellcolor[rgb]{ .906,  .902,  .902}0.01  & \cellcolor[rgb]{ .906,  .902,  .902}0.99  & \cellcolor[rgb]{ .906,  .902,  .902}0.01  & \cellcolor[rgb]{ .906,  .902,  .902}0.01  & \cellcolor[rgb]{ .906,  .902,  .902}0.99  & \cellcolor[rgb]{ .906,  .902,  .902}0.04  & \cellcolor[rgb]{ .906,  .902,  .902}0.01  & \cellcolor[rgb]{ .906,  .902,  .902}0.99  & \cellcolor[rgb]{ .906,  .902,  .902}0.04  & \cellcolor[rgb]{ .906,  .902,  .902}0.18  & \cellcolor[rgb]{ .906,  .902,  .902}0.82  & \cellcolor[rgb]{ .906,  .902,  .902}0.05  & \cellcolor[rgb]{ .906,  .902,  .902}0.00  & \cellcolor[rgb]{ .906,  .902,  .902}1.00  & \cellcolor[rgb]{ .906,  .902,  .902}0.02  & \cellcolor[rgb]{ .906,  .902,  .902}0.03  & \cellcolor[rgb]{ .906,  .902,  .902}0.97  & \cellcolor[rgb]{ .906,  .902,  .902}0.03  & \cellcolor[rgb]{ .906,  .902,  .902}0.03  & \cellcolor[rgb]{ .906,  .902,  .902}0.97  \\
\cmidrule{2-27}          & \multirow{3}[2]{*}{MobileViT-xs} & STRIP & 0.22  & 0.40  & 0.60  & 0.24  & 0.14  & 0.86  & 0.04  & 0.03  & 0.97  & 0.93  & 0.04  & 0.96  & 0.32  & 0.45  & 0.55  & 0.62  & 0.07  & 0.93  & 0.36  & 0.21  & 0.79  & 0.39  & 0.19  & 0.81  \\
          &       & FreqDetector & 0.23  & 0.50  & 0.50  & 0.01  & 0.02  & 0.98  & 0.10  & 0.17  & 0.83  & 0.00  & 0.00  & 1.00  & 0.48  & 0.32  & 0.68  & 0.18  & 0.43  & 0.57  & 0.05  & 0.08  & 0.92  & 0.15  & 0.22  & 0.78  \\
          &       & \cellcolor[rgb]{ .906,  .902,  .902}Ours & \cellcolor[rgb]{ .906,  .902,  .902}0.04  & \cellcolor[rgb]{ .906,  .902,  .902}0.00  & \cellcolor[rgb]{ .906,  .902,  .902}1.00  & \cellcolor[rgb]{ .906,  .902,  .902}0.05  & \cellcolor[rgb]{ .906,  .902,  .902}0.00  & \cellcolor[rgb]{ .906,  .902,  .902}1.00  & \cellcolor[rgb]{ .906,  .902,  .902}0.03  & \cellcolor[rgb]{ .906,  .902,  .902}0.02  & \cellcolor[rgb]{ .906,  .902,  .902}0.98  & \cellcolor[rgb]{ .906,  .902,  .902}0.03  & \cellcolor[rgb]{ .906,  .902,  .902}0.00  & \cellcolor[rgb]{ .906,  .902,  .902}1.00  & \cellcolor[rgb]{ .906,  .902,  .902}0.04  & \cellcolor[rgb]{ .906,  .902,  .902}0.01  & \cellcolor[rgb]{ .906,  .902,  .902}0.99  & \cellcolor[rgb]{ .906,  .902,  .902}0.11  & \cellcolor[rgb]{ .906,  .902,  .902}0.14  & \cellcolor[rgb]{ .906,  .902,  .902}0.86  & \cellcolor[rgb]{ .906,  .902,  .902}0.03  & \cellcolor[rgb]{ .906,  .902,  .902}0.02  & \cellcolor[rgb]{ .906,  .902,  .902}0.98  & \cellcolor[rgb]{ .906,  .902,  .902}0.05  & \cellcolor[rgb]{ .906,  .902,  .902}0.03  & \cellcolor[rgb]{ .906,  .902,  .902}0.97  \\
    \midrule
    \multirow{6}[4]{*}{ImageNet200} & \multirow{3}[2]{*}{WideResNet101-2} & STRIP & 0.02  & 0.04  & 0.96  & 0.13  & 0.12  & 0.88  & -     & -     & -     & 0.11  & 0.15  & 0.85  & 0.28  & 0.37  & 0.63  & 0.03  & 0.02  & 0.98  & 0.31  & 0.37  & 0.63  & 0.15  & 0.18  & 0.82  \\
          &       & FreqDetector & 0.40  & 0.56  & 0.44  & 0.01  & 0.02  & 0.98  & -     & -     & -     & 0.00  & 0.00  & 1.00  & 0.11  & 0.88  & 0.12  & 0.04  & 0.08  & 0.92  & 0.02  & 0.04  & 0.96  & 0.09  & 0.27  & 0.73  \\
          &       & \cellcolor[rgb]{ .906,  .902,  .902}Ours & \cellcolor[rgb]{ .906,  .902,  .902}0.04  & \cellcolor[rgb]{ .906,  .902,  .902}0.00  & \cellcolor[rgb]{ .906,  .902,  .902}1.00  & \cellcolor[rgb]{ .906,  .902,  .902}0.04  & \cellcolor[rgb]{ .906,  .902,  .902}0.00  & \cellcolor[rgb]{ .906,  .902,  .902}1.00  & \cellcolor[rgb]{ .906,  .902,  .902}- & \cellcolor[rgb]{ .906,  .902,  .902}- & \cellcolor[rgb]{ .906,  .902,  .902}- & \cellcolor[rgb]{ .906,  .902,  .902}0.02  & \cellcolor[rgb]{ .906,  .902,  .902}0.00  & \cellcolor[rgb]{ .906,  .902,  .902}1.00  & \cellcolor[rgb]{ .906,  .902,  .902}0.04  & \cellcolor[rgb]{ .906,  .902,  .902}0.02  & \cellcolor[rgb]{ .906,  .902,  .902}0.98  & \cellcolor[rgb]{ .906,  .902,  .902}0.00  & \cellcolor[rgb]{ .906,  .902,  .902}0.00  & \cellcolor[rgb]{ .906,  .902,  .902}1.00  & \cellcolor[rgb]{ .906,  .902,  .902}0.03  & \cellcolor[rgb]{ .906,  .902,  .902}0.02  & \cellcolor[rgb]{ .906,  .902,  .902}0.98  & \cellcolor[rgb]{ .906,  .902,  .902}0.03  & \cellcolor[rgb]{ .906,  .902,  .902}0.01  & \cellcolor[rgb]{ .906,  .902,  .902}0.99  \\
\cmidrule{2-27}          & \multirow{3}[2]{*}{SwinT-Base} & STRIP & 0.08  & 0.19  & 0.81  & 0.10  & 0.10  & 0.90  & -     & -     & -     & 0.98  & 0.01  & 0.99  & 0.34  & 0.53  & 0.47  & 0.64  & 0.03  & 0.97  & 0.38  & 0.23  & 0.77  & 0.42  & 0.18  & 0.82  \\
          &       & FreqDetector & 0.40  & 0.56  & 0.44  & 0.01  & 0.02  & 0.98  & -     & -     & -     & 0.00  & 0.00  & 1.00  & 0.19  & 0.78  & 0.22  & 0.04  & 0.07  & 0.93  & 0.02  & 0.04  & 0.96  & 0.11  & 0.25  & 0.75  \\
          &       & \cellcolor[rgb]{ .906,  .902,  .902}Ours & \cellcolor[rgb]{ .906,  .902,  .902}0.04  & \cellcolor[rgb]{ .906,  .902,  .902}0.00  & \cellcolor[rgb]{ .906,  .902,  .902}1.00  & \cellcolor[rgb]{ .906,  .902,  .902}0.02  & \cellcolor[rgb]{ .906,  .902,  .902}0.01  & \cellcolor[rgb]{ .906,  .902,  .902}0.99  & \cellcolor[rgb]{ .906,  .902,  .902}- & \cellcolor[rgb]{ .906,  .902,  .902}- & \cellcolor[rgb]{ .906,  .902,  .902}- & \cellcolor[rgb]{ .906,  .902,  .902}0.04  & \cellcolor[rgb]{ .906,  .902,  .902}0.12  & \cellcolor[rgb]{ .906,  .902,  .902}0.88  & \cellcolor[rgb]{ .906,  .902,  .902}0.04  & \cellcolor[rgb]{ .906,  .902,  .902}0.01  & \cellcolor[rgb]{ .906,  .902,  .902}0.99  & \cellcolor[rgb]{ .906,  .902,  .902}0.01  & \cellcolor[rgb]{ .906,  .902,  .902}0.00  & \cellcolor[rgb]{ .906,  .902,  .902}1.00  & \cellcolor[rgb]{ .906,  .902,  .902}0.07  & \cellcolor[rgb]{ .906,  .902,  .902}0.05  & \cellcolor[rgb]{ .906,  .902,  .902}0.95  & \cellcolor[rgb]{ .906,  .902,  .902}0.04  & \cellcolor[rgb]{ .906,  .902,  .902}0.03  & \cellcolor[rgb]{ .906,  .902,  .902}0.97  \\
    \bottomrule
    \end{tabular}%
    \begin{tablenotes}
    \Large
    \item[*] LF is computationally infeasible on ImageNet200.
    \vspace{-0.4cm}
    \end{tablenotes}
    \end{threeparttable}
    }
\caption{The evaluation results on different attacks, datasets, and backbones. We observe that the results in additional metrics (FAR, FRR, and \textit{Backdoored Data Rejection Rate} (BDR)) with optimal thresholds are aligned with the conclusions in the paper.} 
\label{main_table}%
\end{table*}%

\begin{table*}[!t]
\setlength{\belowcaptionskip}{-0cm}
  \resizebox{0.98\textwidth}{!}
  {
    \centering
    \begin{tabular}{c|ccc|ccc|ccc|ccc|ccc|ccc}
    \toprule
    Avg, of & \multicolumn{3}{c|}{CIFAR10} & \multicolumn{3}{c|}{GTSRB} & \multicolumn{3}{c|}{CIFAR100} & \multicolumn{3}{c|}{Tiny-ImageNet} & \multicolumn{3}{c|}{ImageNet200} & \multicolumn{3}{c}{AVG} \\
\cmidrule{2-19}     ACC  & $E=1$ & $E=10$ & $E=50$ & $E=1$ & $E=10$ & $E=50$ & $E=1$ & $E=10$ & $E=50$ & $E=1$ & $E=10$ & $E=50$ & $E=1$ & $E=10$ & $E=50$ & $E=1$ & $E=10$ & $E=50$ \\
    \midrule
    CNNs  & 0.7766  & 0.7802  & 0.8078  & 0.8931  & 0.9011  & 0.8968  & 0.8850  & 0.8730  & 0.8823  & 0.9618  & 0.9618  & 0.9618  & 0.9773  & 0.9773  & 0.9773  & 0.8987  & 0.8987  & 0.9052  \\
    ViTs  & 0.7066  & 0.8000  & 0.7801  & 0.8349  & 0.8779  & 0.8687  & 0.9097  & 0.8998  & 0.8957  & 0.9492  & 0.9336  & 0.9377  & 0.9145  & 0.9639  & 0.9639  & 0.8630  & 0.8950  & 0.8892  \\
    \bottomrule
    \end{tabular}%
    }
    \vspace{-0.2cm}
\caption{The accuracy of TeCo in the settings where defenders can estimate the thresholds based on $n$ clean images}
\label{estimated_table}%
\vspace{-0.1cm}
\end{table*}%
\begin{table*}[!t]
\setlength{\belowcaptionskip}{-0cm}
  \resizebox{0.98\textwidth}{!}
  {
    \centering
    \begin{tabular}{c|ccc|ccc|ccc|ccc|ccc|ccc}
    \toprule
    Avg, of & \multicolumn{3}{c|}{CIFAR10} & \multicolumn{3}{c|}{GTSRB} & \multicolumn{3}{c|}{CIFAR100} & \multicolumn{3}{c|}{Tiny-ImageNet} & \multicolumn{3}{c|}{ImageNet200} & \multicolumn{3}{c}{AVG} \\
\cmidrule{2-19}     ACC  & STRIP & FreqDetector & Ours  & STRIP & FreqDetector & Ours  & STRIP & FreqDetector & Ours  & STRIP & FreqDetector & Ours  & STRIP & FreqDetector & Ours  & STRIP & FreqDetector & Ours \\
    \midrule
    CNNs  & 0.6188  & 0.8245  & \textbf{0.8939 } & 0.7008  & 0.7395  & \textbf{0.8899 } & 0.5868  & \textbf{0.8053 } & 0.7434  & 0.6735  & \textbf{0.8200 } & 0.8101  & 0.8135  & 0.8135  & \textbf{0.9760 } & 0.6787  & 0.8006  & \textbf{0.8627}  \\
    ViTs  & 0.5917  & \textbf{0.8233 } & 0.7665  & 0.4988  & \textbf{0.7687 } & 0.7668  & 0.6349  & \textbf{0.8066 } & 0.7381  & 0.6896  & 0.7920  & \textbf{0.8778 } & 0.6735  & 0.8153  & \textbf{0.9639 } & 0.6177  & 0.8012  & \textbf{0.8226}  \\
    \bottomrule
    \end{tabular}%
    }
    \vspace{-0.2cm}
\caption{The accuracy of TeCo and two baselines in the settings where only one statistical threshold can be set to detect all attacks}
\label{static_table}%
\vspace{-0.1cm}
\end{table*}%
\begin{table*}[!t]
\setlength{\belowcaptionskip}{-0cm}
  \resizebox{0.98\textwidth}{!}
  {
    \centering
    \begin{tabular}{ccccccccccccccccccc}
    \toprule
        Avg, of  & \multicolumn{3}{c}{CIFAR10} & \multicolumn{3}{c}{GTSRB} & \multicolumn{3}{c}{CIFAR100} & \multicolumn{3}{c}{Tiny-ImageNet} & \multicolumn{3}{c}{ImageNet200} & \multicolumn{3}{c}{AVG} \\
    \cmidrule{2-19}     ACC  & $\gamma=0$ & $\gamma=0.5$ & $\gamma=1$ & $\gamma=0$ & $\gamma=0.5$ & $\gamma=1$ & $\gamma=0$ & $\gamma=0.5$ & $\gamma=1$ & $\gamma=0$ & $\gamma=0.5$ & $\gamma=1$ & $\gamma=0$ & $\gamma=0.5$ & $\gamma=1$ & $\gamma=0$ & $\gamma=0.5$ & $\gamma=1$ \\
    \midrule
    CNNs  & 0.6672  & 0.7824  & 0.8521  & 0.6604  & 0.7802  & 0.9242  & 0.8111  & 0.8367  & 0.7735  & 0.7351  & 0.7933  & 0.6504  & 0.6125  & 0.7309  & 0.7613  & 0.6973  & 0.7847  & 0.7923  \\
    ViTs  & 0.6130  & 0.7345  & 0.8018  & 0.6132  & 0.7236  & 0.8366  & 0.7816  & 0.8440  & 0.7778  & 0.7460  & 0.8590  & 0.7569  & 0.6313  & 0.7435  & 0.7610  & 0.6770  & 0.7809  & 0.7868  \\
    \bottomrule
    \end{tabular}%
    }
    \vspace{-0.2cm}
\caption{The accuracy of TeCo in the settings where only one empirical threshold can be set to detect all attacks}
\label{empirical_table}%
\vspace{-0.1cm}
\end{table*}%
\begin{table*}[!t]
\setlength{\belowcaptionskip}{-0cm}
  \resizebox{0.98\textwidth}{!}
  {
    \centering
    \begin{tabular}{cc|cc|cc|cc|cc|cc|cc|cc}
    \toprule
    \multicolumn{2}{c|}{Group} & \multicolumn{2}{c|}{$\mathcal{G}_1$} & \multicolumn{2}{c|}{$\mathcal{G}_2$} & \multicolumn{2}{c|}{$\mathcal{G}_3$} & \multicolumn{2}{c|}{$\mathcal{G}_4$} & \multicolumn{2}{c|}{$\mathcal{G}_{1+2}$} & \multicolumn{2}{c|}{$\mathcal{G}_{1+3}$} & \multicolumn{2}{c}{$\mathcal{G}_{1+4}$} \\
    \midrule
    \multicolumn{2}{c|}{Metric} & AUROC & F1 score & AUROC & F1 score & AUROC & F1 score & AUROC & F1 score & AUROC & F1 score & AUROC & F1 score & AUROC & F1 score \\
    \multicolumn{2}{c|}{Avg. of AVG(↑)} & 0.780  & 0.782  & 0.661  & 0.677  & 0.637  & 0.669  & 0.536  & 0.543  & 0.906  & 0.902  & 0.907  & 0.908  & 0.900  & 0.901  \\
    \multicolumn{2}{c|}{Avg. of STD(↓)} & 0.184  & 0.178  & 0.171  & 0.156  & 0.226  & 0.172  & 0.081  & 0.081  & 0.082  & 0.084  & 0.104  & 0.084  & 0.095  & 0.092  \\
    \midrule
    \multicolumn{2}{c|}{$\mathcal{G}_{2+3}$} & \multicolumn{2}{c|}{$\mathcal{G}_{2+4}$} & \multicolumn{2}{c|}{$\mathcal{G}_{3+4}$} & \multicolumn{2}{c|}{$\overline{\mathcal{G}_1}$} & \multicolumn{2}{c|}{$\overline{\mathcal{G}_2}$} & \multicolumn{2}{c|}{$\overline{\mathcal{G}_3}$} & \multicolumn{2}{c|}{$\overline{\mathcal{G}_4}$} & \multicolumn{2}{c}{ALL} \\
    \midrule
    AUROC & F1 score & AUROC & F1 score & AUROC & F1 score & AUROC & F1 score & AUROC & F1 score & AUROC & F1 score & AUROC & F1 score & AUROC & F1 score \\
    0.756  & 0.760  & 0.734  & 0.743  & 0.708  & 0.713  & 0.771  & 0.775  & 0.935  & 0.931  & 0.923  & 0.920  & 0.938  & 0.929  & \textbf{0.945 } & \textbf{0.940 } \\
    0.183  & 0.170  & 0.187  & 0.171  & 0.199  & 0.183  & 0.189  & 0.175  & 0.050  & 0.052  & 0.060  & 0.064  & 0.042  & 0.041  & \textbf{0.035 } & \textbf{0.034 } \\
    \bottomrule
    \end{tabular}%
    }
    \vspace{-0.2cm}
\caption{The performance of TeCo based on different image corruption sets. Results are averaged from different attacks, datasets, and backbones.}
\label{corruption_set}%
\vspace{-0.1cm}
\end{table*}%
\begin{table*}[!t]
\setlength{\belowcaptionskip}{-0cm}
  \resizebox{0.98\textwidth}{!}
  {
    \centering
    \begin{tabular}{c|cccccccccc}
    \toprule
    Measure & \multicolumn{2}{c}{Standard Deviation} & \multicolumn{2}{c}{Range} & \multicolumn{2}{c}{Mean‌ ‌Deviation} & \multicolumn{2}{c}{Coefficient of Variation} & \multicolumn{2}{c}{Quartile Deviation} \\
    \midrule
    Metric & AUROC & F1 score & AUROC & F1 score & AUROC & F1 score & AUROC & F1 score & AUROC & F1 score \\
    Avg. of AVG(↑) & 0.944  & \textbf{0.939 } & 0.912  & 0.906  & \textbf{0.945 } & \textbf{0.939 } & 0.895  & 0.906  & 0.708  & 0.710  \\
    Avg. of STD(↓) & \textbf{0.035 } & \textbf{0.034 } & 0.068  & 0.069  & \textbf{0.035 } & \textbf{0.034 } & 0.075  & 0.062  & 0.186  & 0.180  \\
    \bottomrule
    \end{tabular}%
    }
    \vspace{-0.2cm}
\caption{The performance of TeCo based on different measures of variation. Results are averaged from different attacks, datasets, and backbones.}
\label{deviation_method}%
\vspace{-0.3cm}
\end{table*}%
\begin{table}[!t]
\setlength{\belowcaptionskip}{-0cm}
  \resizebox{0.48\textwidth}{!}
  {
    \centering
    \begin{tabular}{ccc}
    \toprule
    Group & Type  & Corruptions \\
    \midrule
    $\mathcal{G}_1$  & Noise & Gaussian Noise, Shot Noise, Impulse Noise \\
    $\mathcal{G}_2$  & Blur  & Defocus Blur, Glass Blur, Motion Blur, Zoom Blur \\
    $\mathcal{G}_3$  & Nature & Snow, Frost, Fog, Brightness \\
    $\mathcal{G}_4$  & Digital & Contrast, Elastic Transform, Pixelate, Jpeg Compression \\
    \bottomrule
    \end{tabular}%
    }
    \vspace{-0.2cm}
\caption{Images corruptions in different groups.}
\label{groups}%
\vspace{-0.5cm}
\end{table}%

\section{Additional Experiments and Discussions}\label{addition}
\subsection{Thresholds}
Since TeCo maps the input image $x$ to a linearly separable space and defenders make judgments by a threshold $\gamma$, questions are how we can get this threshold and what is the influence of threshold for our method. We investigate these questions in three scenarios: (1) calculating appropriate thresholds from clean data (this seems to have broken the ``no need for extra data" characteristic of TeCo, we will discuss this later.). (2) setting single statistical and static threshold for all potential attacks. (3) setting empirical threshold directly. We evaluate the effectiveness of TeCo in these three scenarios. We use ACC as the evaluation metric, which is calculated by:
\begin{small}
\begin{equation}\label{eq:static_th}
\begin{aligned}
ACC = \frac{TP+TN}{TP+TN+FP+FN}.
\end{aligned}
\end{equation}
\end{small}ACC is enough to estimate the effectiveness because the number of test clean images and the number of test trigger samples are very close according to Tab.~\ref{datasets}.

\textbf{Effectiveness on estimated threshold. }In this setting, we assume defenders can estimate thresholds based on a small set of test clean samples. The estimated threshold is calculated by:
\begin{small}
\begin{equation}\label{eq:static_th}
\begin{aligned}
\gamma_{est} = \frac{1}{E}\sum^E_{e=1} Dev(\mathcal L_e),
\end{aligned}
\end{equation}
\end{small}where $E$ is the number of clean images used to estimate thresholds, $Dev$ is the deviation measurement method, and $L_e$ is the recorded severity list for $e$-th clean image. Tab~\ref{estimated_table} shows the average performance of TeCo in different attacks, datasets, and backbones. These results indicate that TeCo can achieve high effectiveness with a small number of clean data.

\textbf{Effectiveness on statistical and static threshold. }
In some real-world scenarios, the defenders can only set a single prior threshold for all possible attacks. Thus, we investigate the performance of TeCo and two baselines in the static thresholds settings, where only one threshold can be set to detect all the backdoor attacks. The statistical and static threshold is calculated by:
\begin{small}
\begin{equation}\label{eq:static_th}
\begin{aligned}
\overline{\gamma} = \frac{1}{M}\sum^M_{m=1} \arg\max\limits_{\gamma\in\Gamma} \frac{2 \times (\text{precision}_\gamma \times \text{recall}_\gamma)}{(\text{precision}_\gamma + \text{recall}_\gamma)},
\end{aligned}
\end{equation}
\end{small}where $M$ is the number of backdoor attacks. Tab~\ref{static_table} shows the accuracy of the detection methods. TeCo achieves the best effectiveness in $50\%$ settings and the best average effectiveness. These results suggest TeCo can be a practical solution and have performance comparable with the SOTA method which works on looser conditions. 

\textbf{Effectiveness on the empirical threshold. }The most simple way to set the thresholds is to choose common values directly. Tab.~\ref{empirical_table} shows the average performance of TeCo in different attacks, datasets, and backbones when an empirical threshold is given. The results suggest that by empirically setting threshold $=1$, TeCo can still get an average ACC $\approx0.79$, which is a satisfying performance compared with the results in Tab.~\ref{estimated_table}. Since the standard deviation is always larger than or equal to $0$, it is easy to choose $1$ as the threshold without estimating on clean data.

\underline{\textbf{Statement of No Need for Extra Data}}

In our paper, we claim that the proposed method TeCo is independent of extra clean data. However, someone may get confused because theoretically TeCo still needs clean data to get the most appropriate thresholds. We emphasize TeCo's ``no need for extra data" characteristic from two aspects: On the one hand, compared with black-box TTSD methods, TeCo is free of extra data in the linearly separable space mapping process, which is clearly different from existing methods. For example, STRIP superimposes various clean images on the suspicious samples, and FreqDetector needs clean data to serve as the training set of the trigger sample detector. These methods cannot map the input data into a linearly separable space without clean data. On the other hand, other TTSD methods need clean data to gain appropriate thresholds, which seems similar to TeCo. However, TeCo is still different from them because according to Tab.~\ref{estimated_table} and Tab.~\ref{empirical_table}, we can directly set a threshold for TeCo (for example, set $\gamma=1$) without estimating on clean data and enjoy similar performance compared with estimated thresholds. Take Beatrix~\cite{ma2022beatrix} as a counterexample, Beatrix is a white-box TTSD method that needs clean data to get appropriate thresholds. According to the paper, the appropriate threshold of Beatrix on CIFAR10 is about $0.02$, however for GTSRB, the appropriate threshold is about $1.0$, which means the best thresholds of Beatrix among different datasets are quite different, making it hard to set empirical thresholds. 

In a nutshell, for most TTSD methods, the need for extra data is a necessary condition for their effectiveness. On the contrary, extra clean data is \underline{neither sufficient nor necessary} for TeCo. And this is why we can claim TeCo has no need for extra data.

\subsection{Ablation Studies of Image Corruption Set}
We investigate the influence of image corruption set by dividing the involved $15$ image corruptions into $4$ groups, as shown in Tab.~\ref{groups}. Tab.~\ref{corruption_set} presents the performance of TeCo based on different combinations of image corruption groups. The results suggest that only relying on a single type of corruption is not sufficient to get high effectiveness, which is a misunderstanding in related works as we mentioned in our paper. With more corruptions being considered, the performance of TeCo grows correspondingly, indicating that the diversity of image corruptions is an important factor for gaining effectiveness and stability across different attacks and datasets.

\subsection{Ablation Studies of Variation Metrics}
We investigate the influence of the deviation measurement method $Dev$ by introducing four more metrics: Range~\footnote{\url{https://en.wikipedia.org/wiki/Range_(statistics)}}, Mean Deviation~\footnote{\url{https://en.wikipedia.org/wiki/Average_absolute_deviation}}, Coefficient of Variation~\footnote{\url{https://en.wikipedia.org/wiki/Coefficient_of_variation}}, and Quartile Deviation~\footnote{\url{https://en.wikipedia.org/wiki/Interquartile_range}}. Tab.~\ref{deviation_method} presents the performance of TeCo based on different deviation measurement methods.

\subsection{Discussion of Outliers}
There are some interesting results about baselines. Since the Low-frequency (LF) attack is designed to avoid FreqDetector~\cite{zeng2021rethinking}, FreqDetector should have low effectiveness against this attack. However, we implement them following the official codes and find that if we let FreqDetector work in a binary classification manner and make judgments based on thresholds, it will perform well on LF attack. So we believe it is not unfair to involve LF attack and FreqDetector simultaneously in our experiments. To prove the implementation correctness of FreqDetector, we share the performance of FreqDetector on its original mode in Tab.~\ref{detector}, which indicates that LF attack does avoid its detection if FreqDetector works on its original mode. In addition, Wanet can avoid detection, which is aligned with the results in our paper.

Another interesting phenomenon is the success of STRIP against Input-aware and LIRA attacks on SwinTransformer-base/ImageNet200, while STRIP fails on other datasets and backbones. We re-run these experiments by setting different random seeds to ensure the stability of the results. As shown in Tab.~\ref{strip_run}, the results from different random seeds are similar, indicating that the performance of STRIP is somehow influenced by the choice of datasets and backbones. 
\begin{table}[t!]
\resizebox{0.47\textwidth}{!}
{
    \centering
    \begin{tabular}{ccccccccc}
    \toprule
    \multirow{3}[6]{*}{Runs} & \multicolumn{4}{c}{Input-aware} & \multicolumn{4}{c}{LIRA} \\
\cmidrule{2-9}          & \multicolumn{2}{c}{SwinT-B} & \multicolumn{2}{c}{WideResNet} & \multicolumn{2}{c}{SwinT-B} & \multicolumn{2}{c}{WideResNet} \\
\cmidrule{2-9}          & AUROC & F1 score & AUROC & F1 score & AUROC & F1 score & AUROC & F1 score \\
    \midrule
    Run\#1 & 0.936 & 0.86  & 0.423 & 0.504 & 0.994 & 0.975 & 0.696 & 0.645 \\
    Run\#1 & 0.939 & 0.864 & 0.383 & 0.502 & 0.994 & 0.976 & 0.684 & 0.667 \\
    \bottomrule
    \end{tabular}%
}
\caption{The additional random runs of STRIP on ImageNet200}
\label{strip_run}%
\end{table}%
\begin{table}[t!]
\resizebox{0.47\textwidth}{!}
{
    \centering
    \begin{tabular}{ccccccccc}
    \toprule
    Dataset & Accuracy & Badnets & Blended & LF    & Input-aware & Wanet & LIRA  & SSBA \\
    \midrule
    \multirow{2}[2]{*}{CIFAR10} & Trigger Samples & 93.38  & 79.40  & 54.41  & 99.89  & 3.74  & 64.84  & 59.29  \\
          & Clean Images & 96.91  & 96.91  & 96.91  & 96.91  & 96.91  & 96.91  & 96.91  \\
    \midrule
    \multirow{2}[2]{*}{GTSRB} & Trigger Samples & 91.62  & 97.05  & 82.01  & 73.18  & 3.98  & 10.50  & 12.97  \\
          & Clean Images & 94.28  & 94.28  & 94.28  & 94.28  & 94.28  & 94.28  & 94.28  \\
    \midrule
    \multirow{2}[2]{*}{CIFAR100} & Trigger Samples & 87.98  & 77.54  & 58.47  & 98.58  & 1.87  & 81.42  & 54.02  \\
          & Clean Images & 96.17  & 96.17  & 96.17  & 96.17  & 96.17  & 96.17  & 96.17  \\
    \midrule
    \multirow{2}[2]{*}{Tiny-ImageNet} & Trigger Samples & 17.94  & 99.32  & 54.80  & 99.94  & 4.75  & 82.64  & 92.23  \\
          & Clean Images & 98.41  & 98.41  & 98.41  & 98.41  & 98.41  & 98.41  & 98.41  \\
    \midrule
    \multirow{2}[2]{*}{ImageNet200} & Trigger Samples & 1.39  & 98.03  & -     & 100.00  & 0.72  & 86.90  & 94.67  \\
          & Clean Images & 99.07  & 0.00  & -     & 37.20  & 99.07  & 37.20  & 99.07  \\
    \bottomrule
    \end{tabular}%
}
\caption{The effectiveness of FreqDetector in its original mode}
\label{detector}%
\end{table}%

\subsection{Additional Experiments}
\begin{figure}[ht]
    \begin{subfigure}{0.22\textwidth}
        \includegraphics[width=\textwidth]{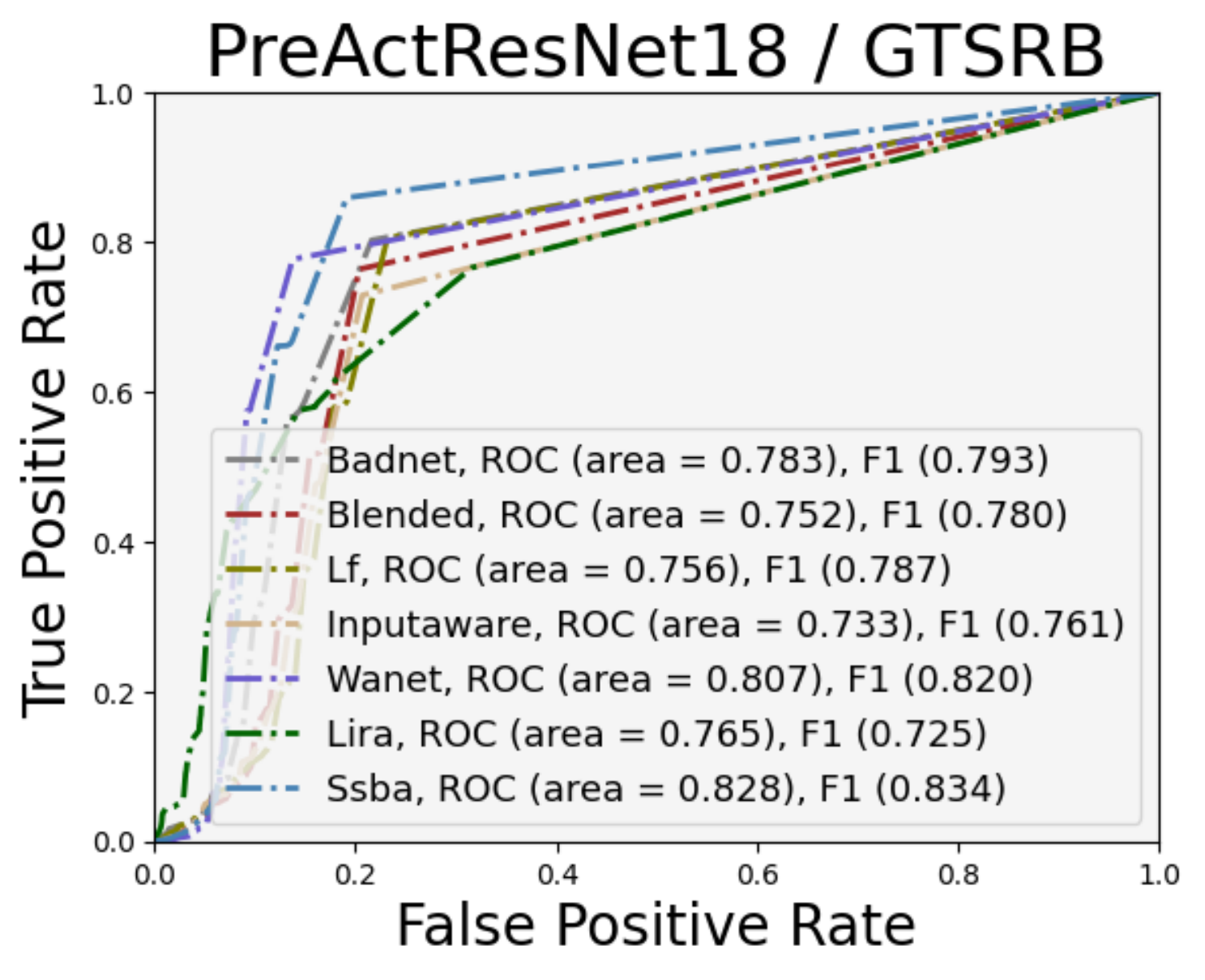} 
        \caption{Ours}
        \label{Fig.all_ours}
    \end{subfigure}
    \begin{subfigure}{0.22\textwidth}
        \includegraphics[width=\textwidth]{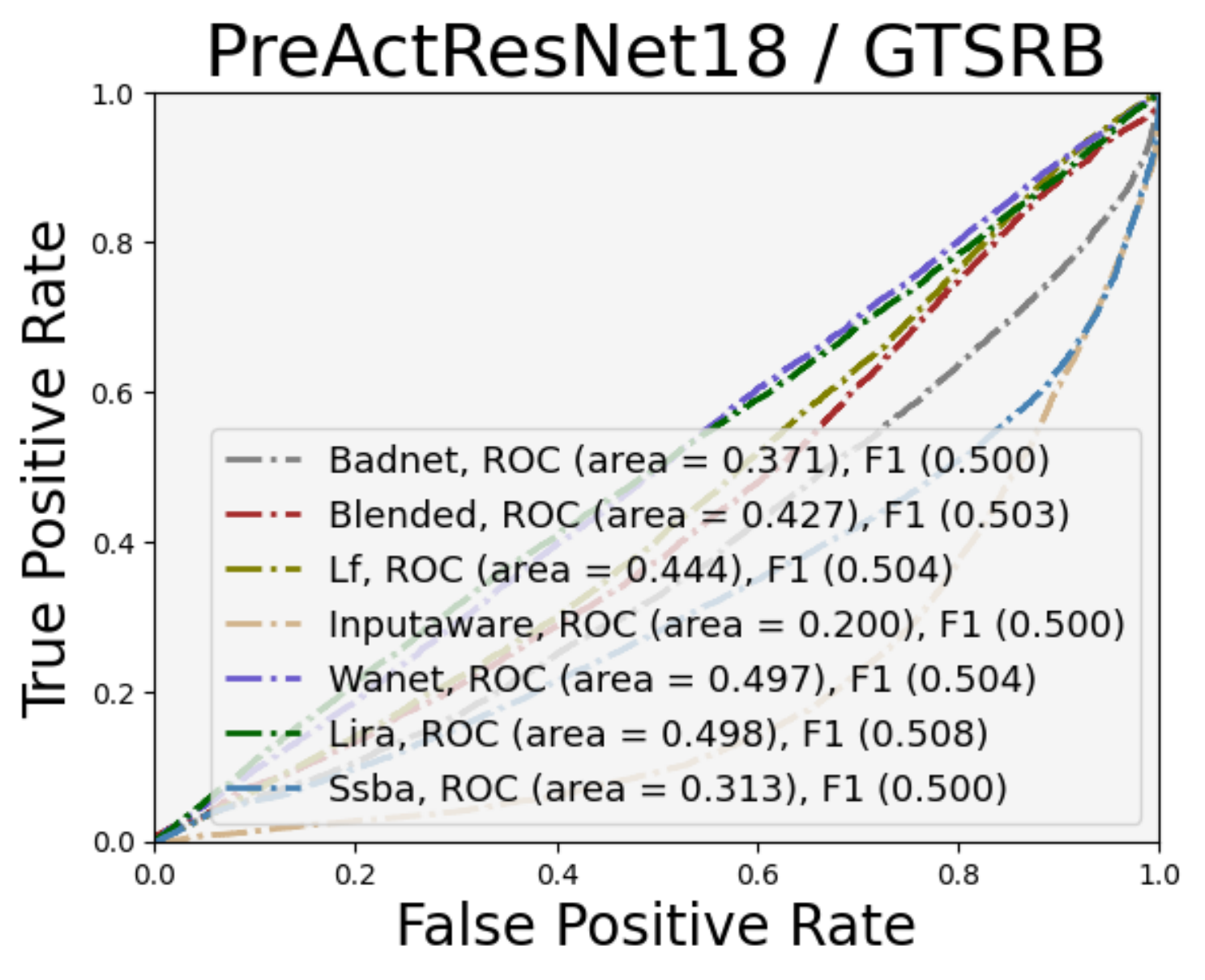} 
        \caption{STRIP}
        \label{Fig.all_strip}
    \end{subfigure}
\caption{ROC of detecting all-to-all attacks}
\vspace{-0.4cm}
\label{Fig.mix}
\end{figure}

\textbf{Effectiveness against all-to-all attacks. }
In practice, a backdoor-infected model may have multiple labels embedded with Trojans, \ie, the multiple classes scenario. Here, we consider the worst multiple classes scenario (all-to-all attack) where every class in the victim model is attacked and the backdoor trigger can cause a specific transition of trigger samples' labels (e.g., turning $y_i$ to $y_i+3$). We investigate TeCo and two baselines against the all-to-all attack on PreActResNet18/GTSRB\footnote{We find all-to-all backdoor attacks are not stable enough on other datasets and cause difficulties to do evaluations.}. As depicted in Fig.~\ref{Fig.mix}(d)-(e), the performance of our method drops about $20\%$ in this scenario but still maintains stability across different attacks. STRIP has lost its performance totally and even makes contrary predictions. Since FreqDetector makes judgments only based on the images, it maintains its performance as the same as which Tab.~\ref{main_table} shows. Fortunately, TeCo is still comparable with FreqDetector in this worst-case setting as demonstrated in Tab.~\ref{targets_table}. 

\begin{table}[htbp]
\setlength{\belowcaptionskip}{-0.3cm}
   \resizebox{0.49\textwidth}{!}
   {
    \centering
    \begin{tabular}{ccccc}
    \toprule
    Method & Avg. of AUROC & Avg. of F1 score & Std. of AUROC & Std. of F1 score \\
    \midrule
    STRIP & 0.3930  & 0.5026  & 0.0997  & 0.0027  \\
    FreqDetector & 0.7911  & 0.7671  & 0.2235  & 0.2027  \\
    \rowcolor[rgb]{ .906,  .902,  .902} Ours  & 0.7749  & 0.7856  & 0.0306  & 0.0336  \\
    \bottomrule
    \end{tabular}%
    }
\caption{Quantization results of detecting all-to-all attacks}
\label{targets_table}%
\end{table}%

\textbf{Smaller triggers.} The results show that TeCo is effective and even more strong against the backdoor attack with smaller triggers.
\begin{table}[h]
\setlength{\belowcaptionskip}{-0.3cm}
\centering
   \resizebox{0.49\textwidth}{!}
   {
    \begin{tabular}{cccccccc}
    \toprule
    Size  & \% of image & AUROC(↑) & F1 score(↑) & ACC(↑) & FAR(↓) & FRR(↓) & BDR(↑) \\
    \midrule
    7*7   & 0.10  & 0.9963  & 0.9969  & 99.69  & 0.60  & 0.01  & 99.99  \\
    14*14 & 0.39  & 0.9973  & 0.9974  & 99.74  & 0.49  & 0.02  & 99.98  \\
    21*21 & 0.88  & 0.9784  & 0.9782  & 97.82  & 4.12  & 0.23  & 99.77  \\
    \bottomrule
    \end{tabular}%
    }
\caption{ImageNet200 / SwinT-Base}
\label{size}%
\end{table}%

\textbf{Transferability to unseen attacks.} The results show in Tab.~\ref{transfer} that TeCo is effective against unseen attacks after optimizing a threshold using a known attack.

\begin{table}[h]
\vspace{-0.3cm}
\setlength{\belowcaptionskip}{-0.3cm}
\centering
   \resizebox{0.49\textwidth}{!}
   {
  \centering
  \begin{tabular}{cccccccccc}
    \toprule
          &       & Badnets & Blended & LF    & Input-Aware & Wanet & LIRA  & SSBA  & AVG \\
    \midrule
    \multirow{4}[2]{*}{Badnets} & ACC   & -     & 0.9451  & 0.9339  & 0.8647  & 0.8559  & 0.9322  & 0.8475  & 0.8966  \\
          & FAR   & -     & 0.1017  & 0.1155  & 0.2096  & 0.1987  & 0.1183  & 0.1641  & 0.1513  \\
          & FRR   & -     & 0.0029  & 0.0112  & 0.0528  & 0.0833  & 0.0117  & 0.1396  & 0.0502  \\
          & BDR   & -     & 0.9971  & 0.9888  & 0.9472  & 0.9167  & 0.9883  & 0.8604  & 0.9498  \\
    \midrule
    \multirow{4}[2]{*}{Blended} & ACC   & 0.9169  & -     & 0.9366  & 0.8703  & 0.8629  & 0.9215  & 0.8324  & 0.8901  \\
          & FAR   & 0.1119  & -     & 0.1099  & 0.1987  & 0.1849  & 0.1138  & 0.1590  & 0.1464  \\
          & FRR   & 0.0511  & -     & 0.0117  & 0.0531  & 0.0840  & 0.0392  & 0.1771  & 0.0694  \\
          & BDR   & 0.9489  & -     & 0.9883  & 0.9469  & 0.9160  & 0.9608  & 0.8229  & 0.9306  \\
    \bottomrule
    \end{tabular}%
    }
\vspace{-0.3cm}
\caption{CIFAR10 / PreActResNet18}
\vspace{-0.15cm}
\label{transfer}%
\end{table}%

\textbf{Corruptions as data augmentations.} The results show that the degradation of performance is not large when the backdoor attacks use corruption for data augmentation. 
\begin{table}[h]
\setlength{\belowcaptionskip}{-0.3cm}
\centering
   \resizebox{0.38\textwidth}{!}
   {
    \begin{tabular}{cccccc}
    \toprule
    Augmentation & Metric & Badnets & Blended & SSBA  & AVG \\
    \multirow{3}[2]{*}{Aug, 50\%} & AUROC(↑) & 0.9040  & 0.9038  & 0.8968  & 0.9015  \\
          & F1 score(↑) & 0.8890  & 0.8863  & 0.8922  & 0.8892  \\
          & BDR(↑) & 93.77  & 95.51  & 95.31  & 94.86  \\

    \bottomrule
    \end{tabular}%
    }
\caption{GTSRB / MobileViT-xs}
\label{aug}%
\end{table}%

\textbf{More recent attacks.} We show additional results on detecting Sleeper Agent~\cite{souri2021sleeper}.
\begin{table}[h]
\setlength{\belowcaptionskip}{-0.3cm}
\centering
   \resizebox{0.49\textwidth}{!}
   {
    \begin{tabular}{ccccccc}
    \toprule
    Attack & AUROC(↑) & F1 score(↑) & ACC(↑) & FAR(↓) & FRR(↓) & BDR(↑) \\
    \midrule
    Sleeper & 0.8897  & 0.9325  & 93.25  & 10.80  & 2.70  & 97.30  \\
    \bottomrule
    \end{tabular}%
    }
\caption{CIFAR10 / PreActResNet18}
\label{new_attack}%
\end{table}%

\begin{figure*}[t]
\centering
\setlength{\belowcaptionskip}{-0.1cm}
    \begin{subfigure}{0.49\textwidth}
        \includegraphics[width=\textwidth]{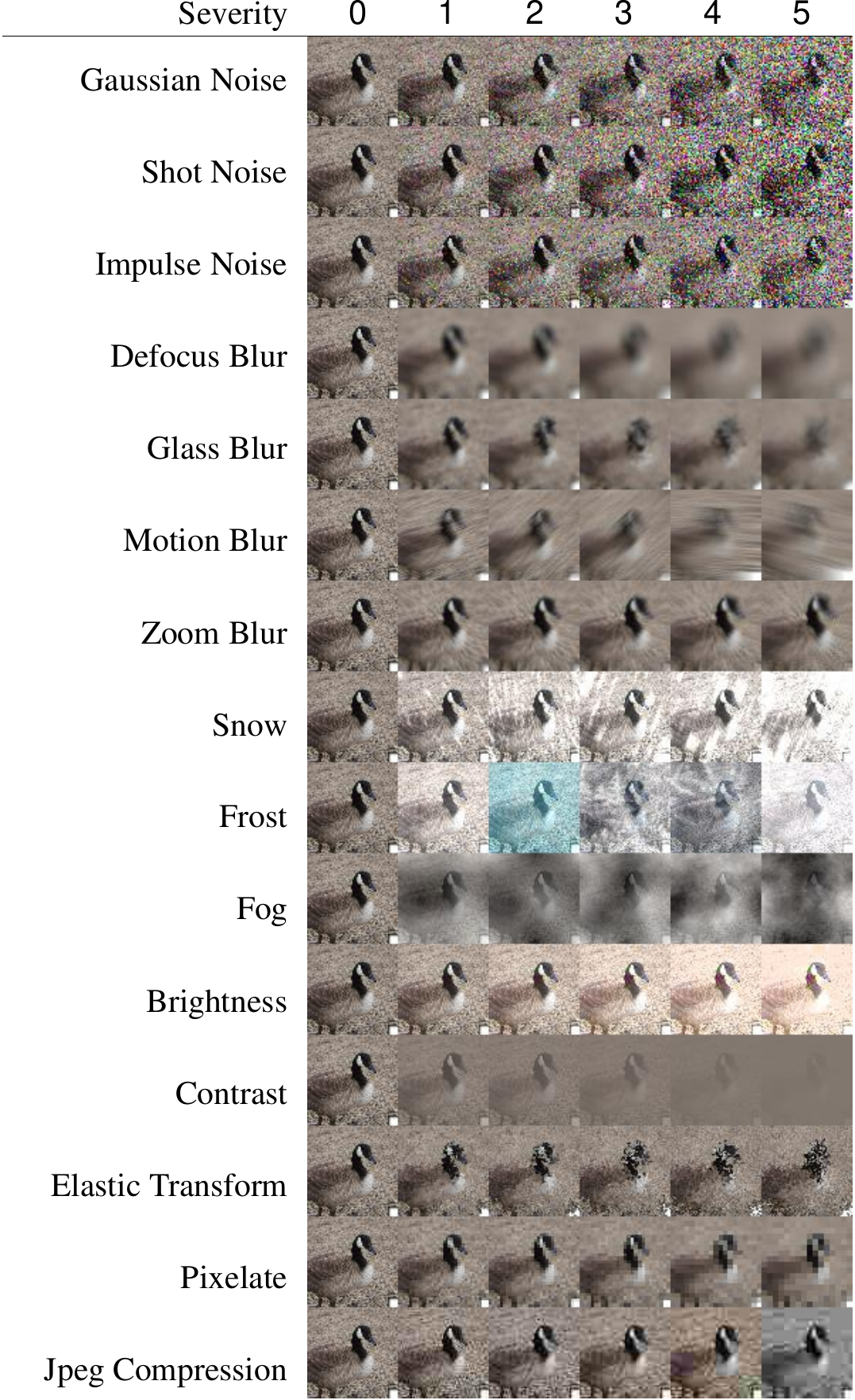} 
        \caption{Badnets / Tiny-ImageNet.}
        \label{asr_heatmap_sec3}
    \end{subfigure}
    \begin{subfigure}{0.49\textwidth}
        \includegraphics[width=\textwidth]{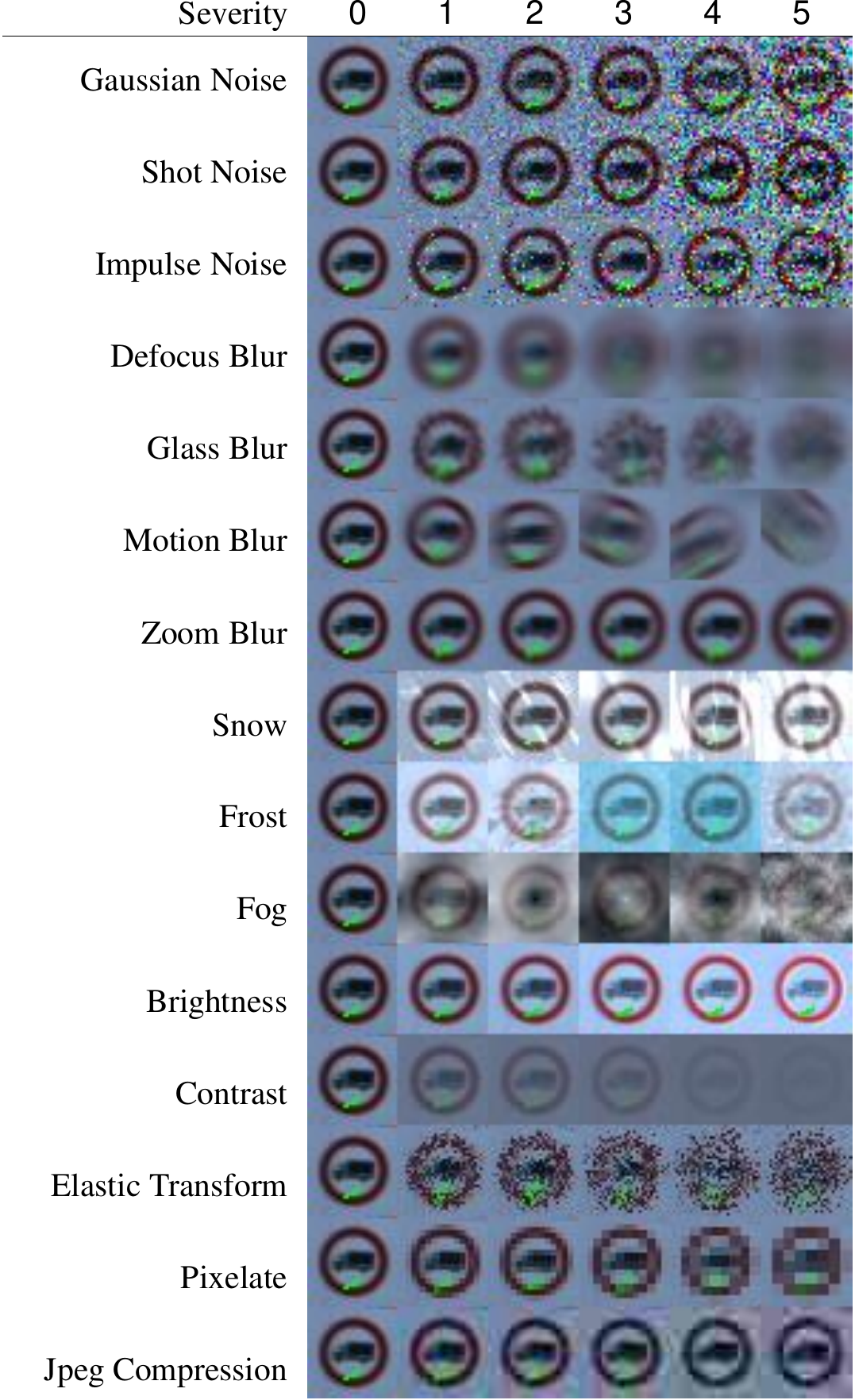}
        \caption{Input-aware / GTSRB.}
        \label{acc_heatmap_sec3}
    \end{subfigure}
\label{vis}
\caption{Visualization of trigger samples and their corrupted versions}
\vspace{-0.3cm}
\end{figure*}

\textbf{Insightful discussion of TeCo.} As we discussed in Sec.6, the explanation of TeCo is very likely to be the dual-target training function of backdoor attacks which leads to the huge bias of victim models. The bias makes victim models focus on the trigger patterns rather than the original information of trigger samples. When the trigger patterns encounter different corruptions, since some corruptions are in texture information while others are in structure information, the trigger will be robust against certain corruptions while not robust against others. And since clean images have more complex texture and structure information compared with trigger patterns which need to be simple and repetitive for causing bias, the clean images will have consistent robustness. In this paper, our main goal is to discover and introduce this phenomenon to the community with comprehensive empirical studies. A formal theoretical study will be our future work.

\textbf{Use TeCo to detect backdoor-infected models.} As we have mentioned in our introduction, TeCo is a \textit{test-time trigger sample detection} (TTSD) method that can seamlessly integrate into existing model diagnosis defenses for defense. In practice, defenders can first use model diagnosis defenses (e.g., AEVA~\cite{DBLP:conf/iclr/GuoLL22}, which also works in hard-label black-box settings) to judge whether the target model is a backdoor model. Then the defenders can use TeCo to detect the trigger samples. On the other hand, TeCo can be used to diagnose the model. Our study shows that for the clean samples on clean models, the FAR of TeCo will be high when applying thresholds calculated from the backdoor model (Avg. FAR$\approx55\%$ on GTSRB/PreActResNet18 clean model). So defenders may feed a batch of clean images into the target model, and calculate the FAR of TeCo to judge whether the target model is a backdoor model.

\end{document}